\documentclass[11pt]{article}
\pdfoutput=1

\usepackage{jheppub}
\usepackage{latexsym}
\usepackage{multirow}
\usepackage{color}
\usepackage[usenames,dvipsnames,table]{xcolor}
\usepackage{graphicx}% Include figure files
\usepackage{epsfig}  % Include figure files
\usepackage{dcolumn}% Align table columns on decimal point
\usepackage{bm}% bold math
\usepackage{dcolumn}% Align table columns on decimal point
\usepackage{textcomp}% Align table columns on decimal point
\usepackage{float}
\usepackage{hypcap}
\usepackage[]{hyperref}
\usepackage{makecell}
\usepackage{multirow}
\usepackage{color}
\usepackage{pifont}
\usepackage{subfigure}
\usepackage{amssymb}
 \usepackage{caption}
\usepackage{array}
\newcolumntype{L}[1]{>{\raggedright\let\newline\\\arraybackslash\hspace{0pt}}m{#1}}
\newcolumntype{C}[1]{>{\centering\let\newline\\\arraybackslash\hspace{0pt}}m{#1}}
\newcolumntype{R}[1]{>{\raggedleft\let\newline\\\arraybackslash\hspace{0pt}}m{#1}}
\newcommand{\la}{\lambda}

\newcommand{\wt}{\widetilde}
\newcommand{\ov}{\overline}
 \usepackage[normalem]{ulem}
\usepackage{color}

\allowdisplaybreaks
  \newcommand{\capdef}{}
  \newcommand{\mycaption}[2][\capdef]{\renewcommand{\capdef}{#2}
       \caption[#1]{{\footnotesize #2}}}
  \newcommand{\be}{\begin{equation}}
   \newcommand{\ee}{\end{equation}}

\hypersetup{
  bookmarks=true,         % show bookmarks bar?
  unicode=false,          % non-Latin characters in Acrobat?s bookmarks
  pdftoolbar=true,        % show Acrobat?s toolbar?
 pdfmenubar=true,        % show Acrobat?s menu?
 pdffitwindow=true,     % window fit to page when opened
 pdfstartview={FitH},    % fits the width of the page to the window
 pdfsubject={Neutrino Oscillations Phenomenology},   % subject of the document
 pdfnewwindow=true,      % links in new window
 pdfcreator={RevTeX},
 colorlinks=true,       % false: boxed links; true: colored links
 linkcolor=red,          % color of internal links
 citecolor=blue,        % color of links to bibliography
 filecolor=black,      % color of file links
 urlcolor=blue,           % color of external links
  }

\title{Sub-TeV Quintuplet Minimal Dark Matter with Left-Right Symmetry}

\author[a,b]{Sanjib Kumar Agarwalla,}
\author[a,b]{Kirtiman Ghosh,}
\author[c]{Ayon Patra}

\affiliation[a]{Institute of Physics, Sachivalaya Marg, Sainik School Post, Bhubaneswar 751005, India}
\affiliation[b]{Homi Bhabha National Institute, Training School Complex, Anushakti Nagar, Mumbai 400085, India}
\affiliation[c]{Centre for High Energy Physics, Indian Institute of Science, Bangalore - 560012, India}

\emailAdd{sanjib@iopb.res.in}
\emailAdd{kirtiman.ghosh@iopb.res.in}
\emailAdd{ayon@okstate.edu}

 \abstract{A detailed study of a fermionic quintuplet dark matter in a left-right symmetric 
scenario is performed in this article. The minimal quintuplet dark matter model 
is highly constrained from the WMAP dark matter relic density (RD) data. 
To elevate this constraint, an extra singlet scalar is introduced.
It introduces a host of new annihilation and co-annihilation channels for the 
dark matter, allowing even sub-TeV masses. The phenomenology of this singlet 
scalar is studied in detail in the context of the Large Hadron Collider (LHC) experiment. 
The production and decay of this singlet scalar at the LHC give rise to interesting resonant 
di-Higgs or diphoton final states. We also constrain the RD allowed parameter space 
of this model in light of the ATLAS bounds on the resonant di-Higgs and diphoton 
cross-sections.
 }
\keywords{Beyond Standard Model, Dark Matter, Higgs Physics, Left--Right Symmetry }
\arxivnumber{1803.01670}
\begin{document}
\preprint{IP/BBSR/2018-3}
\maketitle
\flushbottom

%==========================
\section{Introduction and Motivation}
%==========================
  
Dark matter (DM) and its large abundance compared to baryonic matter has been a long standing puzzle without any definite answer as of yet. The Standard Model (SM) of particle physics is bereft of any such particles and hence the need to extend the SM becomes imperative in order to incorporate a DM candidate in the particle spectrum. Numerous approaches have been made to come up with consistent DM models that can explain the experimental observations. A small class of such models is what is referred to as Minimal dark matter (MDM) ~\cite{Cirelli:2005uq,Heeck:2015qra,Ko:2015uma,Garcia-Cely:2015quu%,Dey:2015bur
,Agarwalla:2016rmw,Maru:2017pwl} models. MDM models postulate a new fermionic or bosonic multiplet, an n-tuplets of the $SU(2)$ group. Being color neutral these new multiplets have no strong interactions, and can only weakly interact with other SM particles mainly through gauge interactions. The stability of the quintuplet, on the other hand, is either ensured by some discrete symmetry or they can be accidentally stable. In a scenario where the lightest component of the new multiplet is electrically neutral, it could be a good candidate for the DM. In this work we will study a MDM model where the DM is coming from the neutral component of a quintuplet fermion.

A dark matter coming from $SU(2)_L$ quintuplet has severe limitations. Firstly, only a hypercharge zero quintuplet can evade the strong direct detection limits with all other cases being very highly constrained. All the states in the quintuplet fermion are mass degenerate at the tree level. This degeneracy is lifted (only by few hundred MeV) by radiative corrections. This makes the {collider phenomenology for the} quintuplet extremely challenging. Quintuplets, being charged under the SM gauge group, are produced in pairs at the collider experiments via the gauge interactions. Subsequently, they decay into lightest component of the quintuplet in association with very soft SM leptons and jets. The lightest component of the quintuplet, being the candidate for the DM, remains invisible in the detectors whereas, the final state leptons or jets will be too soft (due to the extremely small mass splitting) to observe at the collider. Attempts have been made to overcome this difficulty by introducing an additional quadruplet scalar \cite{Kumericki:2012bh, Yu:2015pwa} in order to write a dimension-4 decay term for the quintuplet fermion. However, in this case, the dark matter candidate would then be lost or only be there in an extremely fine-tuned region of the parameter space. We thus take an alternate approach to this problem by choosing a left-right symmetric model where the DM is neutral component of an $SU(2)_R$ quintuplet fermion. 

Left-right symmetric (LRS) models~\cite{Mohapatra:1974hk,Senjanovic:1975rk} by themselves are a very well motivated extension of the SM. They are gauge extensions of the SM with the gauge group being $SU(3)_C\times SU(2)_L\times SU(2)_R\times U(1)_{B-L}$. At a fundamental level, LRS models preserve parity (P) symmetry. The spontaneous breaking of the right-handed symmetry at some high scale leads to the observed parity violation at the weak scale. Another important consequence of this fact is that the P-violating terms in the QCD Lagrangian leading to the strong-CP problem \cite{Beg:1978mt,Mohapatra:1978fy,Babu:1989rb,Barr:1991qx,Mohapatra:1995xd,Kuchimanchi:1995rp,Mohapatra:1997su,Babu:2001se,Kuchimanchi:2010xs} are absent in these class of models and hence naturally solving the strong-CP without a global Peccei-Quinn symmetry \cite{Peccei:1977hh}. The gauge structure here compels us to have a right-handed neutrino in the particle spectrum, thus allowing for a small neutrino mass generation through the seesaw mechanism \cite{Minkowski:1977sc,Yanagida:1979as,Sawada:1979dis,Levy:1980ws,VanNieuwenhuizen:1979hm,Mohapatra:1979ia}. 

In this work, we have considered $SU(3)_C\times SU(2)_L\times SU(2)_R\times U(1)_{B-L}$ gauge symmetry and enlarged the fermion spectrum by introducing a vector-like fermion multiplet which is a quintuplet under $SU(2)_R$. The neutral component of the $SU(2)_R$ quintuplet could be good candidate for the dark matter. As the fundamental gauge group in this case does not include $U(1)_Y$, it is only produced after the right-handed symmetry breaking of $SU(2)_R \times U(1)_{B-L} \rightarrow U(1)_Y$. The hypercharge quantum number is thus a derived quantity and allows for many different combinations of charge assignment for the quintuplet to get the zero hypercharge for the DM particle. This model would thus have a much richer phenomenology with many different possibilities for the DM and other particles of the quintuplet. The tree-level masses of the quintuplet particles are still degenerate but the radiative corrections are much larger in this case. The mass degeneracy among the quintuplet fermions of different charges are now produced at the right-handed symmetry breaking scale (heavy right-handed gauge bosons are running in the loops) resulting in much larger mass splitting among them. Thus the production and subsequent decay of the high charge-multiplicity components of the quintuplet produce interesting signatures at the collider experiments without sacrificing the dark matter aspect of the model.

The dark matter candidate in the present scenario is the neutral component of the vector-like $SU(2)_R$ quintuplet fermion. Therefore, the dark matter has gauge interactions with the $SU(2)_R$ gauge bosons namely, the $W_R$ and $Z_R$-bosons. The self-annihilation and co-annihilation of the dark matter mainly proceed through a $Z_R$ and $W_R$ exchange in the $s$-channel, respectively. It is important to note that the lower values for the masses of $W_R$ and $Z_R$ are highly constrained from the LHC data as well as other low energy observables. As a result, the self-annihilation and co-annihilation cross-sections are, in general, small. The dark matter relic density measured by WMAP \cite{Hinshaw:2012aka} and PLANCK \cite{Ade:2015xua} collaborations can only be satisfied for some particular values of DM masses (at around half the $W_R$ and $Z_R$ masses) where  self-annihilation and/or co-annihilation cross-sections are enhanced by the resonant production of  $Z_R$ and/or $W_R$, respectively. It has been shown in Ref.~\cite{Ko:2015uma} that the dark matter relic density can only be satisfied at quite large DM masses (around 4 TeV or higher) if one also accounts for the direct detection constraints. A way to circumvent this problem, as has been discussed later in this paper, is to introduce a singlet scalar which can open up a lot of new annihilation and co-annihilation channels allowing for a correct DM relic density for almost any DM mass while also satisfying the direct detection bounds. We have studied this in details in this paper, focusing on the DM and singlet scalar phenomenology in the model.

{The scalar, being singlet under both $SU(2)_L$ and $SU(2)_R$, has Yukawa coupling only with the vector like quintuplet fermions. The couplings with the other scalars in the model arise from the scalar potential. However, couplings with the SM gauge bosons, namely $W^\pm,~Z$ and photon, arise at one loop level via the loops involving quintuplet fermions. The collider phenomenology of the singlet scalar crucially depends on its loop induced coupling with a pair of photons. In the framework of this model, singlet scalar-photon-photon coupling is  enhanced due to multi-charged quintuplet fermions in the loop. Therefore, at the LHC experiment, statistically significant number of singlet scalar could be produced via photon-photon\footnote{The inclusion of the photon as a parton inside the proton, with an associated parton distribution function (PDF) is required to include next-to-leading order (NLO) QED corrections. Since $\alpha_S^2$ is of the same order of magnitude as $\alpha_{EM}$ and in the era of precision phenomenology at the LHC when the PDFs are already determined upto NNLO in QCD, consistency of calculations require PDFs which are corrected atleast upto NLO QED.} initial state. Depending on the parameters in the scalar potential, the singlet scalar dominantly decays into a pair of SM Higgses or photons, giving rise to interesting di-Higgs or di-photon resonance signatures at the LHC. We have also studied the collider signatures of the singlet scalar and bounds on the singlet scalar masses from the di-Higgs and di-photon resonance searches by the ATLAS collaboration with 36 inverse femtobarn integrated luminosity data of the LHC running at 13 TeV center-of-mass energy.}

The rest of the paper is organized as follows. In Sec.~\ref{model} we have introduced the model. The particle spectrum is listed along with the interactions among the particles. We have computed the gauge boson, fermion and the scalar masses and mixings in this section. In Sec.~\ref{DarkM} we have studied the dark matter phenomenology of the model along with the bounds from direct detection experiments. In Sec.~\ref{singlet} we study the phenomenology of the singlet scalar. Sec.~\ref{bounds} has the collider bounds from the most recent diphoton and di-Higgs results from the Large Hadron Collider (LHC). Finally we conclude in Sec.~\ref{conclusion} with some discussions. 

\section{Quintuplet Minimal Dark Matter Model}
\label{model}

In this work, we consider a 
minimal model for dark matter (DM) in the framework of $SU(3)_C\times SU(2)_L\times SU(2)_R\times U(1)_{B-L}$ gauge symmetry, where $B$ and $L$ are baryon and lepton numbers respectively. Due to the left-right symmetric nature of the model, the chiral fermions are now doublets for both left and right-handed sectors and are given as:
\begin{eqnarray}
&Q_L& \left(3,2, 1, \frac13 \right ) = \!\left (\begin{array}{c}
u\\ d \end{array} \right )_L,~~~~
Q_R\left ( 3,1, 2, \frac13
\right )\!=\!\left (\begin{array}{c}
u\\d \end{array} \right )_R,\nonumber \\ [4pt]
&l_L&\left ( 1,2, 1, -1 \right )=\left (\begin{array}{c}
\nu\\ e\end{array}\right )_L,~~~~
l_R\left ( 1,1, 2, -1 \right )=\left (\begin{array}{c}
\nu \\ e \end{array}\right )_R,
\end{eqnarray}
where the numbers in the bracket corresponds to $SU(3)_C,~ SU(2)_L,~SU(2)_R~{\text{and}}~U(1)_{B-L}$ quantum numbers respectively. The electric charge $Q$ for a particle in this model is given as: $Q=T^3_L+T^3_R+Q_{(B-L)}/2$, where $T^3_{L/R}$ are the third component of the isospin for $SU(2)_{L/R}$. A minimal scalar sector requires a right-handed doublet Higgs boson to break the $SU(2)_R$ symmetry and a bidoublet Higgs field to break the electroweak symmetry and produce the quark and lepton masses along with the CKM mixings. They are given as:
%=========
\begin{eqnarray}
H_R(1,1,2,1)&=&\left (\begin{array}{c}
H_R^+ \\H_R^0 \end{array} \right ),~~~~
\Phi(1,2,2,0)={\left (\begin{array}{cc}
\phi^{0}_1 & \phi^{+}_{2} \\ \phi^{-}_{1} & \phi^{0}_{2} \end{array} \right)}.~~~
\label{scalar}
\end{eqnarray}
%==========

The absence of a triplet scalar in the Higgs sector prevents us from writing a lepton number violating term in the Yukawa Lagrangian and hence a light neutrino mass generation is not possible in this scenario without introducing unnaturally small Yukawa couplings. We thus introduce a singlet fermion N(1,1,1,0) which will help generate light neutrino mass through inverse seesaw mechanism. 

We introduce an additional $SU(2)_R$ vector-like fermion quintuplet
\begin{eqnarray}
\Sigma(1,1,5,X) = ( \chi^{2+X/2},\chi^{1+X/2},\chi^{X/2},\chi^{-1+X/2},\chi^{-2+X/2})^T,
\end{eqnarray}
to accommodate a DM candidate (the neutral component of the quintuplet). Here $X=0,2,4$ are the possible values of the $B-L$ quantum numbers for $\chi$ and the numbers in the superscript are the charges for individual particles of the quintuplet as a function of $X$. Depending on the $B-L$ quantum number of this quintuplet fermion, only a very few specific values of DM mass can account for the correct relic density. The DM direct detection bounds would constrain the allowed DM masses even further.  An easy way to circumvent this would be to introduce a singlet scalar $S(1,1,1,0)$ which opens up a number of new annihilation and co-annihilation channels to satisfy the DM constraints for almost any DM mass depending on the mass of $S$. This singlet scalar can also give rise to interesting di-Higgs and diphoton final state signals which can further be used to constrain the model.

The neutral component of the $H_R$ acquires a non-zero vacuum expectation value (VEV) denoted by $\left< H^0_R \right> = v_R$ to break the right-handed symmetry with $SU(2)_R \times U(1)_{B-L}$ breaking into $U(1)_Y$. The heavy gauge boson masses are thus naturally generated at this scale. The electroweak (EW) symmetry breaking and the fermion masses and mixings, on the other hand, are generated by the  neutral components of $\Phi$ field once they acquire a non-zero VEV given by $\left< \phi_{1}^0 \right> = v_1$ and $\left< \phi^0_{2} \right> = v_{2}$. For simplicity, we will assume the VEV of the $S$-field to be zero.\footnote{Even if $S$ were to get a non-zero VEV, this could be easily absorbed into other dimension-full parameters of the model.}

The gauge bosons of $SU(2)_L$, $SU(2)_R$, and $U(1)_{B-L}$ mix among themselves to give four massive ($W_R,~Z_R$, and the SM $W$ and $Z$-boson) and one massless (the SM photon) gauge bosons. We denote the left-handed (right-handed) triplet gauge state as $W_L^i (W_R^i)$ while the $B-L$ gauge boson is $B$. The mass-squared matrix for the charged gauge boson $M_W^2$ in the basis $(W_R,W_L)$ and the neutral gauge bosons $M_Z^2$ in the basis $(W_{3R},W_{3L},B)$ is given as
%==============
\begin{equation}
M_W^2 = \frac{1}{2}\begin{bmatrix}
g_R^2\left( v_R^2+v^2 \right)&g_L g_R v_1 v_2 \\ g_L g_R v_1 v_2&g_L^2 v^2
\end{bmatrix},~~~~
M_Z^2 = \frac{1}{2}\begin{bmatrix}
g_R^2\left(v_R^2+v^2\right)&g_L g_R v^2&-g_R g_{B-L} v_R^2 \\ g_L g_R v^2&g_L^2 v^2&0 \\-g_R g_{B-L} v_R^2&0&g_{B-L}^2 v_R^2
\end{bmatrix},
\end{equation}
%===============
where $v^2=v_1^2+v_2^2$ is the EW VEV $\sim$ 174 GeV and $g_R = g_L = 0.653$. This gives the masses of the new right-handed heavy gauge bosons as:
\begin{eqnarray}
M^2_{W_R}&=&\frac{1}{2}g_R^2\left( v_R^2+v^2 \right),~~~~M^2_{Z_R} = \frac{1}{2}\left(g_R^2+g^2_{B-L}\right) \left[ v_R^2+\frac{g_R^2 v^2}{\left(g_R^2+g^2_{B-L}\right)} \right],
\end{eqnarray}
while the left-handed $W$ and $Z$ boson masses are the same as in the SM 
with the effective hypercharge gauge coupling given as $g_Y = (g_R g_{B-L})/(g_R^2+g_{B-L}^2)^{1/2}$. 

The relevant couplings of the gauge bosons with $\chi$ are given as:
%===============
\begin{eqnarray}
L \supset &-&g_Y s_W Q_{\chi^i} \overline{\chi}^i Z^\mu \gamma_\mu \chi^i + e Q_{\chi^i} \overline{\chi}^i A^\mu \gamma_\mu \chi^i + \sqrt{g_R^2-g_Y^2}\left[Q_{\chi^i} - \frac{g_R^2 Q_{B-L}}{2 \left(g_R^2-g_Y^2\right)} \right]\overline{\chi}^i Z_R^\mu \gamma_\mu \chi^i\notag \\
&+&(\frac{g_R}{\sqrt{2}} r_Q \ov\chi^{i+1}W^\mu_R\gamma_\mu\chi^i + h.c.). ~~~~
\label{eq:gaudm}
\end{eqnarray}
%===============
Here $r_Q = \sqrt{(3+Q_{\chi_i}-Q_{B-L}/2)(2-Q_{\chi_i}+Q_{B-L}/2)}$, $\chi^i$ are the particles in the quintuplet $\Sigma$ with $Q_{\chi^i}$ being the corresponding electric charge and $s_W = \sin{\theta_W}$ where $\theta_W$ is the Weinberg angle. These couplings are particularly important from the perspective of DM phenomenology as they can lead to self annihilation (through $Z_R$) and co-annihilation (through $W_R$) of the DM particle so as to satisfy the RD bounds at these points. 

The fermion masses are generated from the following Yukawa Lagrangian:
%==============
\begin{eqnarray}
\mathcal{L}_Y&=& \left(Y_q\overline{Q}_{L}\Phi Q_{R}+\widetilde{Y}_q\overline{Q}_{L}\widetilde{\Phi}Q_{R}+Y_l\overline{l}_{L}\Phi l_{R}+\widetilde{Y}_l\overline{l}_{L}\widetilde{\Phi}l_{R} + f_{R}\overline l_{R} \widetilde H_R N + H.C. \right) \notag \\ 
&+& \frac{{\mu_N}}{2} N N + \lambda S \overline{\Sigma} \Sigma + M_\Sigma \overline{\Sigma} \Sigma,
\end{eqnarray}
%===============
where $Y$ and $f$ are the Yukawa couplings and $\widetilde{\Phi}=\tau_2\Phi^\ast\tau_2,\widetilde H_{R} = i \tau_2 H^\ast_{R}$.
The quark and charged lepton masses in this model would then be given as:
%===========
\begin{eqnarray}
M_{u} &=& Y_q v_1+\wt{Y}_q v_2,~~M_d=Y_q v_2+\wt{Y}_q v_1,~~M_l = Y_l v_2+\wt{Y}_l v_1.~~~~
\label{eq:fermass}
\end{eqnarray}
%===========
For simplicity, we will choose a large $\tan{\beta}~(=v_1/v_2)$ limit which requires 
$Y_{33}^q \sim 1$ to explain the top quark mass while $\wt{Y}_{33}^q < 10^{-2}$. The neutrino mass matrix, on the other hand, is a $3\times3$ matrix in the basis $(\nu_L,\nu_R,N)$ given as:
\begin{equation}
M_{\nu}=\begin{bmatrix}
0&m_D&0\\m_D^T&0&f_R v_R\\0&f_R^T v_R&\mu_N
\end{bmatrix},
\end{equation} 
where $m_D = Y_l v_1+\wt{Y}_lv_2$. This is the inverse seesaw mechanism of neutrino mass generation. If we assume that $f_R v_R >>m_D,\mu_N$ the approximate expressions for the neutrino mass eigenvalues (for one generation) are given as 
\begin{equation}
m_{\nu_1} \sim \frac{(f_R^{-1} M_D^T)^T \mu_N (f_R^{-1} M_D^T)}{v_R^2}, ~~~~m_{\nu_{2,3}} \sim f_R v_R.
\end{equation}
So it is easy to get a light neutrino mass by appropriately choosing all the parameters in the neutrino sector.

The most general scalar potential involving the bidoublet field, an $SU(2)_R$ doublet field, and a real singlet field is given by:
%=====================
\begin{eqnarray}
V_{H} &\supset& - \mu_1^2~ \text{Tr}\left[\Phi^{\dagger} \Phi \right] - \mu_2^2~\text{Tr} \left[\widetilde \Phi \Phi^{\dagger} + H.C. \right] - \mu_R^2 H_R^\dagger H_R - \frac{\mu_S^2}{2} S^2  + \lambda_1 \left[ \text{Tr} \left(\Phi^{\dagger} \Phi\right) \right]^2\nonumber \\
&+& \lambda_2 \left[ \left\{ \text{Tr} \left( \wt \Phi \Phi^{\dagger}\right) \right\}^2 +H.C. \right]+ \lambda_3 \text{Tr} \left( \wt \Phi \Phi^{\dagger}\right) \text{Tr} \left( {\wt \Phi}^{\dagger} \Phi\right) + \lambda_4 \text{Tr} \left(\Phi^{\dagger} \Phi \right) \text{Tr} \left[\widetilde \Phi \Phi^{\dagger} + H.C. \right] \notag \\
&+& \alpha_3 \mu_3 S \text{Tr}\left(\Phi^{\dagger} \Phi \right) + \alpha_4 \mu_4 S \text{Tr} \left[\widetilde \Phi \Phi^{\dagger} + H.C. \right] +\alpha_5 \mu_5 S H_R^\dagger H_R + \frac{\beta_1}{2} S^2 \text{Tr}\left(\Phi^{\dagger} \Phi \right) \nonumber\\
&+&  \frac{\beta_2}{2} S^2 \text{Tr} \left[\widetilde \Phi \Phi^{\dagger} + H.C. \right] 
+ \frac{\beta_3}{2} S^2 H_R^\dagger H_R + \rho_1 H_R^\dagger H_R  \text{Tr}\left[\Phi^{\dagger} \Phi \right] + \rho_2 H_R^\dagger H_R \text{Tr} \left[\widetilde \Phi \Phi^{\dagger} + H.C. \right] \nonumber\\
&+& \rho_3 H_R^\dagger \Phi^\dagger \Phi H_R + \lambda_R \left( H_R^\dagger H_R \right) ^2 + A_S \mu_S S^3 + \frac{\lambda_S}{4} S^4.
\label{eq:hpot}
\end{eqnarray} 
%======================
The physical Higgs spectrum consists of four CP-even scalars, one CP-odd pseudoscalar, 
and one charged Higgs boson. Two charged states and two CP-odd states are eaten up by the four massive gauge bosons. Using the Higgs potential given in Eqn.~\ref{eq:hpot} to eliminate $\mu_1^2$, $\mu_2^2$ and $\mu_R^2$ from the minimization conditions, we get the CP-even scalar mass-squared matrix as:
\begin{equation}
\hspace*{-0.2cm}
{\small
{\normalsize{M_h^2}} = \begin{bmatrix}M_{11}&M_{12}&M_{13} & M_{14} \\
M_{12}&M_{22}& M_{23} & M_{24} \\
M_{13}& M_{23}&M_{33}&M_{34}\\
M_{14}&M_{24}&M_{34}&M_{44}
\end{bmatrix}}~~~~
\label{hmass}
\end{equation}
where 
\begin{eqnarray}
\hspace*{-2cm}
M_{11}&=& 4\left\{\la_1 v_1^2+2 \la_4 v_1 v_2 + \left(2 \la_2 + \la_3\right)v_2^2\right\}+\frac{\rho_3 v_2^2 v_R^2}{v_1^2-v_2^2}, \notag\\
M_{12}&=&4\left(\la_1+2\la_2+\la_3\right)v_1v_2+4\la_4\left(v_1^2+v_2^2\right)-\frac{\rho_3 v_1 v_2 v_R^2}{v_1^2-v_2^2}, \notag \\
M_{13}&=& \sqrt{2}\left(\alpha_3 \mu_3 v_1+2 \alpha_4 \mu_4 v_2\right),\notag \\
M_{14}&=& 2 \left(\rho_1 v_1+2\rho_2 v_2\right)v_R , \notag \\
M_{22}&=& 4\left\{\la_1 v_2^2+2 \la_4 v_1 v_2 + \left(2 \la_2 + \la_3\right)v_1^2\right\}+\frac{\rho_3 v_1^2 v_R^2}{v_1^2-v_2^2},\notag\\
M_{23}&=& \sqrt{2}\left(\alpha_3 \mu_3 v_2+2 \alpha_4 \mu_4 v_1\right), \notag \\
M_{24}&=&2 \left\{2\rho_2 v_1+(\rho_1+\rho_3) v_2\right\}v_R, \notag \\
M_{33}&=&\beta_1(v_1^2+v_2^2)+4 \beta_2 v_1 v_2 +\beta_3 v_R^2-\mu_S^2, \notag \\
M_{34}&=&\sqrt{2}\alpha_5 \mu_5 v_R, \notag \\
M_{44}&=&4 \la_R v_R^2.
\end{eqnarray}
Diagonalizing this mass-squared matrix gives four mass eigenstates. Table~\ref{tab:one} gives one benchmark point for a set of parameters which can easily give a light SM-like 125 GeV Higgs boson denoted by $h$, consisting almost entirely of the real part of $\phi_1^0$ field. We also get a 500 GeV scalar denoted by $H_1$, consisting of almost purely the singlet $S$ with negligible mixing with the others. This state is the one most important from the dark matter point of view and it is easy to see here that the mass of this state can be easily increased (decreased) by just decreasing (increasing) the value of $\mu_S^2$ and very slightly modifying value of $\la_1$ accordingly. Such a change does not significantly alter the composition of this $H_1$ eigenstate till about a mass of 200 GeV. Further decreasing the mass of $H_1$ (just by increasing $\mu_S^2$) results in significant mixing of the singlet with the SM-like state and is ruled out from Higgs data. Though one can then alter the other parameters of the model to still keep the mixing low. Two very heavy states $H_2$ and $H_3$ with masses of the order of $v_R$ consisting of real part of $\phi_2^0$ and $H_R^0$ states are also present in the spectrum. The heavy states are required to be heavier than 15 TeV in order to suppress flavor changing neutral currents~\cite{Ecker:1983uh,Mohapatra:1983ae,Pospelov:1996fq,Zhang:2007da,Maiezza:2010ic}. 
This can be easily satisfied in our model by choosing a high value ($>$ 10 TeV) for the 
right-handed symmetry breaking scale, $v_R$. The mass of the pseudo-scalar $A_1$ is given as:
\begin{equation}
M_{A_1}^2 = 4\left(\la_3-2\la_2+\frac{\rho_3 v_R^2}{v_1^2-v_2^2}\right)\left(v_1^2+v_2^2\right),
\end{equation} while the charged Higgs boson $H^{\pm}$ mass is
\begin{equation}
M_{H^\pm}^2 = \rho_3 \frac{ \left(v_1^2-v_2^2\right)^2+v_R^2\left(v_1^2+v_2^2\right)}{v_1^2-v_2^2}.
\end{equation} 
%===========
\begin{table}[ht!]
\centering
\begin{tabular}{||C{1.7cm}|C{1.8cm}|C{8.5cm}||}
\hline
Particle & Mass & Composition \\ \hline
$h$& 125 GeV & $0.999 \phi_{1\mathcal{R}}^0+0.029 \phi_{2\mathcal{R}}^0 - 0.016 S-0.003 H_{R\mathcal{R}}^0 $ \\ \hline
$H_1$ & 500 GeV & $0.016 \phi_{1\mathcal{R}}^0 + 0.0004 \phi_{2\mathcal{R}}^0 + 0.999 S  - 0.001 H_{R\mathcal{R}}^0$ \\ \hline
$H_2$ & 18.36 TeV & $0.019 \phi_{1\mathcal{R}}^0 - 0.590 \phi_{2\mathcal{R}}^0 +0.001 S + 0.807 H_{R\mathcal{R}}^0 $ \\ \hline
$H_3$ & 18.43 TeV & $0.021 \phi_{1\mathcal{R}}^0 - 0.807 \phi_{2\mathcal{R}}^0 - 0.001 S - 0.590 H_{R\mathcal{R}}^0 $ \\ \hline
$A_1$ & 18.41 TeV & $0.029 \phi_{1\mathcal{I}}^0 - 0.999 \phi_{2\mathcal{I}}^0 $ \\ \hline
$H^\pm$ & 18.40 TeV & $0.029 \phi_1^\pm + 0.999 \phi_2^\pm + 0.013 H_R^\pm $ \\ \hline
\end{tabular}
\mycaption{Scalar mass eigenstates for $\la_1$ = 0.146, $\la_2$ = 0.35, $\la_3$ = 0.831, 
$\la_4$ = 0.01, $\alpha_3$ = 0.14, $\alpha_4$ = 0.1, $\alpha_5$ = 0.1, $\mu_3$ = 174 GeV, $\mu_4$ = 174 GeV, 
$\beta_1$ = 0.1, $\beta_2$ = 0.1, $\beta_3$ = 0.004, $\rho_1$ = 0.2, $\rho_2$ = 0.2, 
$\rho_3$ = 2, $\mu_S^2$ = $(655)^2$ ${\text{GeV}}^2$, $v_1$ = 173.9 GeV, 
$v_2$ = 5 GeV, $v_R$ = 13 TeV. Subscript $\mathcal{R}$ and $\mathcal{I}$ stands for the real and imaginary parts of the field respectively.}  
\label{tab:one}  
\end{table}
%===========

\section{Dark Matter Phenomenology}
\label{DarkM}

The motivation for introducing the vector-like quintuplet fermion $\chi(1,1,5,X)$ 
was to obtain a candidate for DM. Since all the components of $\Sigma$ get mass from the same term they are all mass degenerate at tree-level, but radiative corrections remove this degeneracy. Radiative corrections to the masses of the quintuplet fermions will be introduced from the gauge sector and the singlet scalar but since the coupling of the singlet scalar to all the quintuplet particles are the same, it does not introduce any splitting between their masses. 
The mass splitting due to quantum corrections is thus given by,
%============
\begin{eqnarray}
M_{\chi^i}-M_{\chi^0}&=&\frac{g_R^2}{(4\pi)^2}M_{\chi^0} {\bigg{[}} Q_{\chi^i}\left(Q_{\chi^i}-Q_{B-L}\right)f(r_{W_R}) - Q_{\chi^i} \left(\frac{g_Y^2}{g_{B-L}^2}Q_{\chi^i}-Q_{B-L} \right)f(r_{Z_R})\nonumber \\
&-& \left. \frac{g_Y^2}{g_R^2}Q_{\chi^i}^2\left\{s_W^2 f(r_{Z}) + c_W^2 f(r_{\gamma}) \right\} \right],
\label{eq:mass}
\end{eqnarray}
%=============
where $r_X$ = $m_X/M_{\chi^0}$, $f(r) \equiv 2 \int_0^1 dx (1+x) \text{log} \left[ x^2 + (1-x)r^2 \right]$ and $Q_{\chi^i}$ is the electric charge of $\chi^i$. 
%===================================================
\begin{figure}[t!] \includegraphics[width=0.325\linewidth]{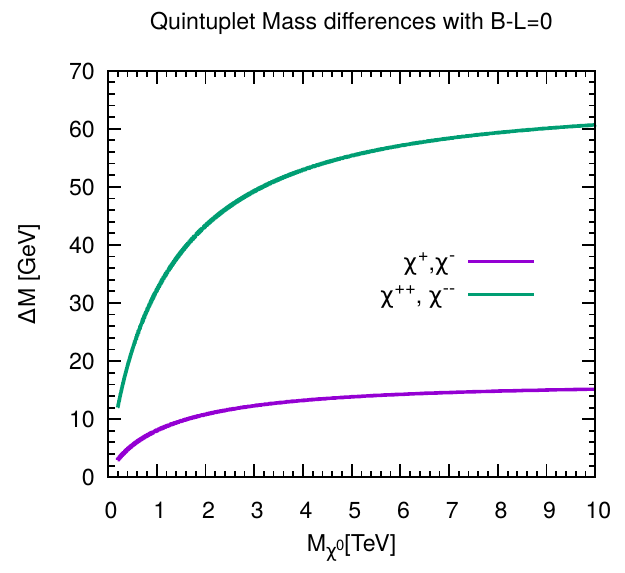} \includegraphics[width=0.325\linewidth]{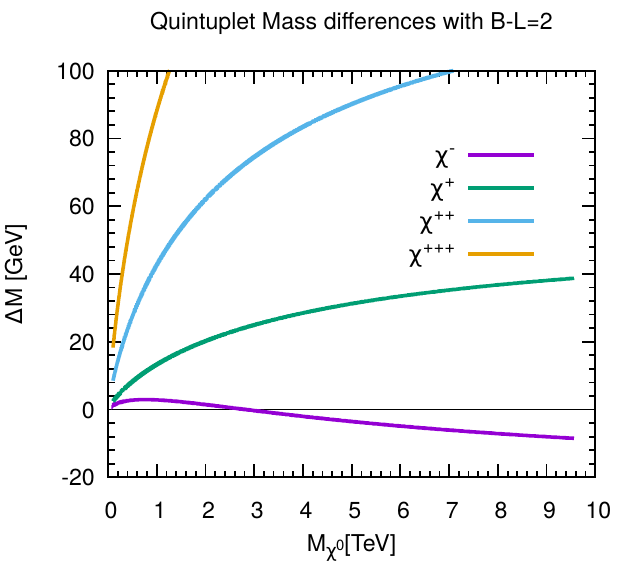} \includegraphics[width=0.325\linewidth]{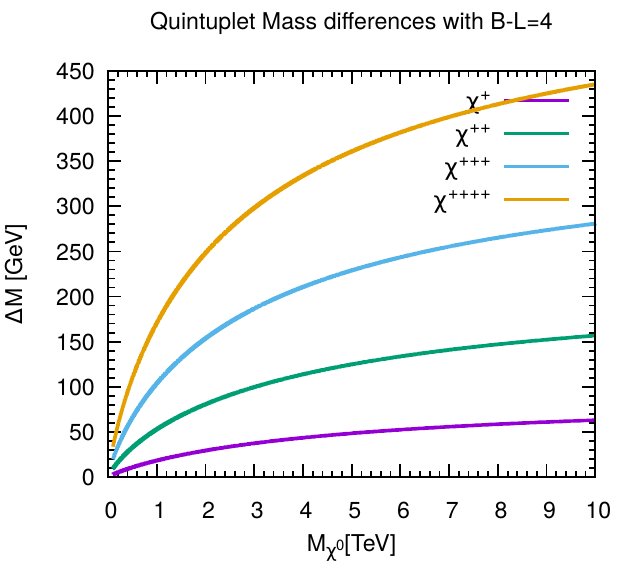} \mycaption{Mass difference between various charged states as a function of the neutral state mass for different B-L cases for $M_{W_R}=6$ TeV and $M_{Z_R}=7.14$ TeV. Note that the y-axis ranges are different in all the panels.} \label{fig:massdiff} \end{figure}

%\begin{figure}[ht!]
%\centering
%\includegraphics[width=2.3in]{MD0.pdf}
%\includegraphics[width=2.3in]{MD2.pdf}
%\includegraphics[width=2.3in]{MD4.pdf}
%\caption{Mass difference between various charged states as a function of the neutral state mass for different B-L cases for $M_{W_R}=6$ TeV and $M_{Z_R}=7.14$ TeV. Note that the y-axis ranges are different in all the panels.}
%\label{fig:massdiff}
%\end{figure}
%===================================================
Fig.~\ref{fig:massdiff} gives a plot of the mass differences between the neutral and the  various charged states for all three cases with $B-L=0,2,4$. For a major portion of the parameter space, the masses of the charged components of the quintuplet get positive contribution from the radiative corrections and hence $\chi^0$ becomes the lightest among the quintuplet fermions\footnote{We will limit ourselves to the region where $\chi^0$ is the lightest and thus discard the region where the negatively charged state becomes the lightest for $B-L=2$ case.}. Thus the lightest component of the quintuplet $\chi^0$ can be a good candidate for dark matter. The stability of $\chi^0$ is automatically ensured by virtue of its gauge quantum numbers. As $\chi^0$ is part of the quintuplet, it can decay to the SM particles only via interactions with dimension-6 or higher operators resulting in a decay width suppressed atleast by a factor of $1/\Lambda^2$. Taking the mass of $\chi^0$ to be at TeV scale, the decay width via dimension-6 operator is of the order of $M^3/\Lambda^2$. This corresponds to a lifetime greater than the age of the universe for $\Lambda \gtrsim 10^{14}$ GeV.

\subsection{Relic Density}
\label{Relic}

%====================================
%=========================================================== 
\begin{figure}[b!]\centering \includegraphics[width=7.7cm]{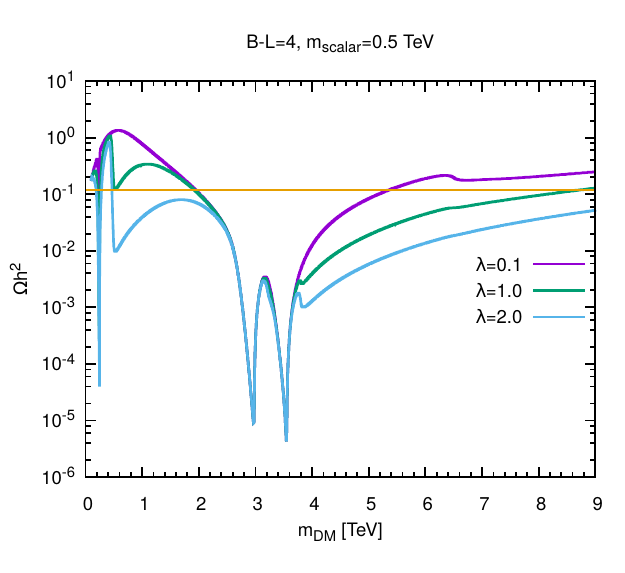} \includegraphics[width=7.7cm]{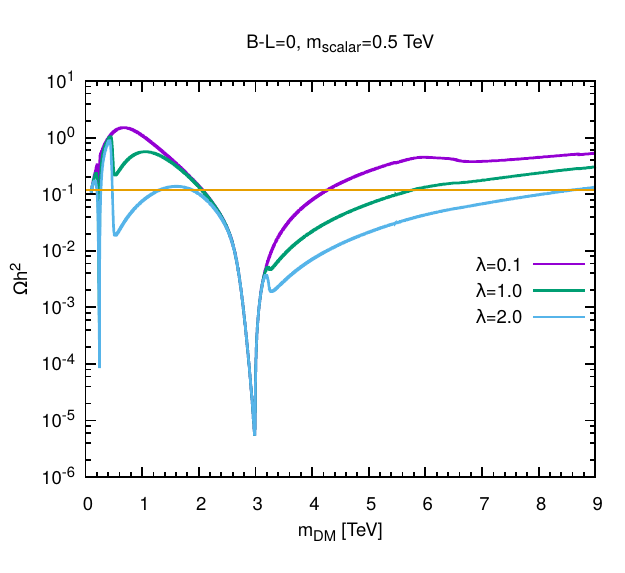} \includegraphics[width=7.7cm]{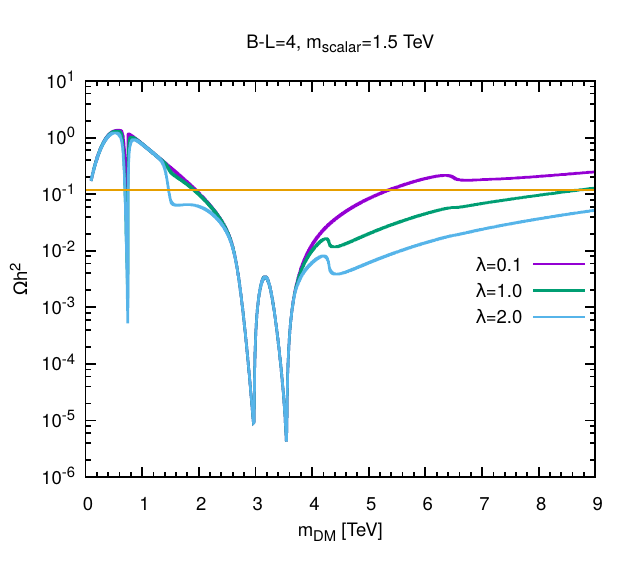} \includegraphics[width=7.7cm]{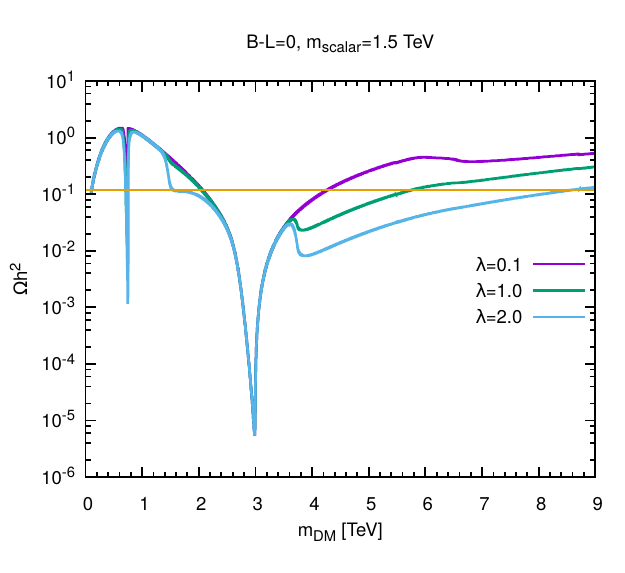} \mycaption{Relic density as a function of the dark matter mass. The left panel is for $B-L = 4$ case, whereas the right panel is for $B-L = 0$ case. The upper (lower) panels are for the scalar mass 0.5 TeV (1.5 TeV). In all the panels, we take $M_{W_R}=6$ TeV, $M_{Z_R}=7.14$ TeV, $\alpha_3\mu_3 = v_{EW}$.} \label{fig:rd4} \end{figure} %=============================================================
%\begin{figure}[t]
%\centering
%\includegraphics[width=3.3in]{bml4al10mh5.pdf}
%\includegraphics[width=3.3in]{bml0al10mh5.pdf}
%\includegraphics[width=3.3in]{bml4al10mh15.pdf}
%\includegraphics[width=3.3in]{bml0al10mh15.pdf}
%\caption{Relic density as a function of the dark matter mass. The left panel is for $B-L = 4$ case, whereas the right panel is for $B-L = 0$ case. The upper (lower) panels are for the scalar mass 0.5 TeV (1.5 TeV). In all the panels, we take $M_{W_R}=6$ TeV, $M_{Z_R}=7.14$ TeV, $\alpha_3\mu_3 = v_{EW}$.}
%\label{fig:rd4}
%\end{figure}
%====================================

The dark matter relic density as a function of the DM mass for $B-L=4$ and $B-L=0$ cases are given in Fig. \ref{fig:rd4}. We have varied the DM mass from 100 GeV to 9 TeV and plotted the relic density for three values of $\la$ corresponding to $\la = 0.1, 1.0, 2.0$ and two fixed values of the scalar masses of 500 GeV and 1.5 TeV respectively while keeping $\alpha_3 \mu_3 = v_{EW}$. The other important numbers required to fully understand the plots are $M_{W_R}=6$ TeV, $M_{Z_R}=7.14$ TeV. Considering first the $B-L=4$ case, it is easy to see that there are five dips in the plot with three of them being very sharp while two others being shallower. These are the regions where a sudden enhancement in the cross-section of either annihilation of two DM particles or co-annihilation of a DM with a singly charged $\chi^{\pm}$ giving rise to a sudden decrease in the relic density. The three regions with sharp fall-off corresponds to three s-channel processes while the two flatter ones correspond to two t-channel processes\footnote{s and t are Mandelstam variables given as $s=(p_1+p_2)^2$ and $t=(p_1-p_3)^2$, where $p_i$ denotes the 4-momentum of particle $i$ for a process $1+2\rightarrow3+4$.}. The first dip at $M_{DM}=M_{scalar}/2$ corresponds to the s-channel process where two DM particles annihilate through an $H_1$ into SM particles. This process reaches its resonance at a DM mass of half of the scalar mass and the sharp fall is because of the s-channel process. Careful analysis of the plots will show that the dip in the left plot is actually deeper than the right one. This is because the total decay width of the 500 GeV scalar particle is smaller than the 1.5 TeV case resulting in a larger resonant cross-section and hence a deeper valley in the left plot.

The second dip is at $M_{DM}=M_{H_1}$ and corresponds to the t-channel process of two DM particles annihilating into two $H_1$ bosons. An interesting thing to note is at larger values of $\la$ the relic density can easily satisfy the experimentally observed value while for smaller $\la$ values, the relic density is greater than the experimental limits. This is because the annihilation cross-section $\sigma_{\chi^0 \chi^0 \rightarrow h_1 h_1} \sim \la^4$ and only for large values of $\la$ will lead to a large enough annihilation for the required decrease in relic density. If we take $\la\rightarrow 0$ then this dip will disappear all together. Another consequence of being a t-channel process is that in the limit $t \rightarrow 0$ the cross-section $\sigma \sim M_{H_1}^{-2}$ and hence larger (smaller) the scalar mass, smaller (larger) the cross-section. This is the reason why the plot on the right with $M_{H_1}=1.5$ TeV has a much smaller decrease in the relic density at this point compared to the left plot with $M_{H_1}=500$ GeV.

The third fall-off corresponds to co-annihilation of the DM particle through a $W_R^\pm$ ($\chi^0~\chi^\pm \rightarrow W_R^\pm \rightarrow SM$) resonance while the fourth is DM annihilation through a $Z_R$ resonance ($\chi^0~\chi^0 \rightarrow Z_R \rightarrow SM$). It is easy to see that the dips are exactly at $M_{W_R^\pm}/2$ and $M_{Z_R}/2$ respectively. The sharp fall-off again is an indication that both are s-channel processes.  

The fifth dip corresponds to t-channel annihilation process of $\chi^0~\chi^0 \rightarrow Z_R~H_1$ and is exactly at a DM mass equal to half of the combined masses of $Z_R$ and $H_1$ bosons. The annihilation cross-section $\sigma \sim \la^2$ in this case and hence this dip will again disappear in the limit $\la\rightarrow0$. Actually there is another t-channel co-annihilation process corresponding to $\chi^0~\chi^\pm \rightarrow W_R^\pm~H_1$ which should have had a dip around a DM mass of $M_{W_R^\pm}+M_{H_1}$ but that is masked by the dip corresponding to the $Z_R$--mediated annihilation channel. 

%======================================================================
%\begin{figure}[ht!]
%\centering
%\includegraphics[width=3.3in]{bml0al10mh5.pdf}
%\includegraphics[width=3.3in]{bml0mh1500.pdf}
%%\includegraphics[width=2.3in]{MD4.pdf}
%\caption{{Relic density plots for B-L=0 case with $M_{W_R}=6$ TeV, $M_{Z_R}=7.14$ TeV.}}
%\label{fig:rd0}
%\end{figure}
%======================================================================

The relic density plot for the $B-L=0$ case is given in the right panels of Fig.~\ref{fig:rd4}. 
This plot has only four dips, the ones involving $Z_R$ are absent here. This is because there is no $\chi^0 \chi^0 Z_R$ vertex as can be obtained from eqn.~\ref{eq:gaudm} by putting $Q_{\chi_i}=0$ along with $Q_{B-L}=0$. Similar to the previous case, the first dip corresponds to the resonance of s-channel annihilation process mediated by $H_1$. The larger (smaller) total decay width in the heavier (lighter) $H_1$ mass case leads to a shallower (deeper) dip like before. The second dip corresponds to the t-channel annihilation process $\chi^0~\chi^0\rightarrow H_1~H_1$. The third dip is the s-channel co-annihilation mediated by a $W_R^\pm$ boson. The fourth dip corresponds to the t-channel co-annihilation of $\chi^0~\chi^\pm \rightarrow W_R^\pm ~H_1$ which was not visible in the previous case. It is clearly visible here as the $Z_R$ boson couplings with the DM particle is absent. This cross-section is again proportional to $\la^2$ and hence the dip decreases with $\la$ and eventually vanishes as $\la\rightarrow 0$.

%===================================
\begin{figure}[!htbp]
\centering
\includegraphics[width=7.7cm]{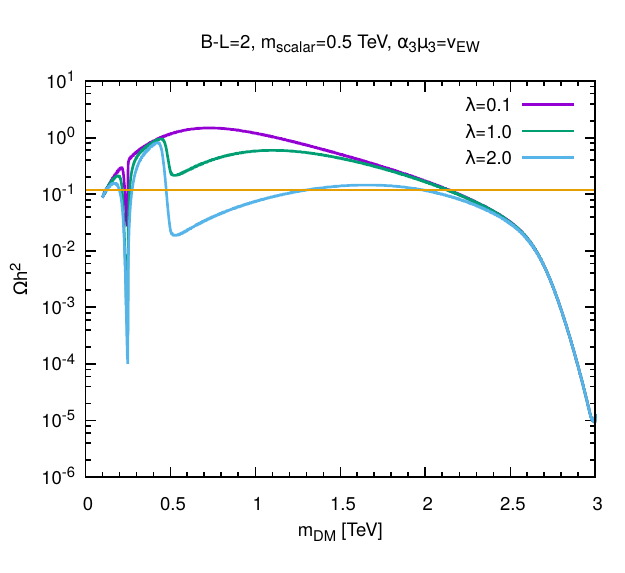}
\includegraphics[width=7.7cm]{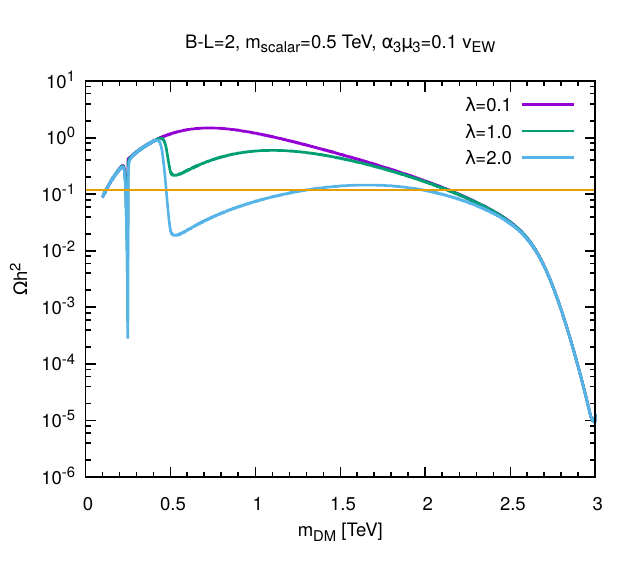}
\mycaption{Relic density as a function of the dark matter mass for B-L=2 case with 
$M_{W_R}=6$ TeV, $M_{Z_R}=7.14$ TeV.}
\label{fig:rd2}
\end{figure}
%===================================

The relic density plot for the $B-L=2$ case is given in Fig.~\ref{fig:rd2}. Here the DM mass is only taken till around 3 TeV as after that the negatively charged component of the quintuplet becomes the lightest and stable and hence, ruled out. The plot is very similar to the previous ones. The three dips visible here are due to s-channel $H_1$ mediated annihilation, the t-channel annihilation of two DM particles into two $H_1$s and the s-channel $W_R^\pm$ mediated co-annihilation processes respectively. 
We have included two plots here for a DM mass of 500 GeV for two different values of $\alpha_3 \mu_3$ being $v_{EW}$ and 0.1$v_{EW}$ respectively. It is easy to see that the only difference between the two plots are in width of the s-channel annihilation region. The lower value of $\alpha_3 \mu_3$ leads to a much narrower region with the relic density being satisfied by two points which are very close in DM mass while for the larger coupling there are two quite distinct values of DM mass possible. This is because at the resonance point, a smaller value of $\alpha_3 \mu_3$ will lead to a smaller annihilation cross-section resulting in a narrower and shallower dip in the relic density plot. Similarly for the other two cases ($B-L$ = 0, 4), the effect of this trilinear coupling would only be seen in the scalar mediated annihilation channel as that is the only relevant process involving this coupling.

The introduction of the scalar singlet $S$ has a huge influence on the allowed dark matter masses which can satisfy the observed relic density. As has been discussed earlier, majority of dips in the dark matter relic density plots would disappear in the absence of this singlet scalar. In fact if $S$ is removed from the spectrum there would be no possible dark matter satisfying the observed relic density for the $B-L=2$ case with a $W_R$ mass of 2 TeV \cite{Ko:2015uma}. The introduction of a singlet even in this constrained parameter space could provide at least two points for correct DM relic density for a small enough singlet scalar mass. We would thus like to examine what happens if we keep the singlet boson mass as a free parameter while also varying $\la$. As has been discussed in Sec.~\ref{model} the singlet--like Higgs boson mass can be easily changed by just varying the value of $\mu_S^2$, hence it is quite natural that the singlet mass is not a fixed quantity but a variable in this analysis. 

%====================================================
\begin{figure}[!htbp]
\centering
\includegraphics[width=7.6cm]{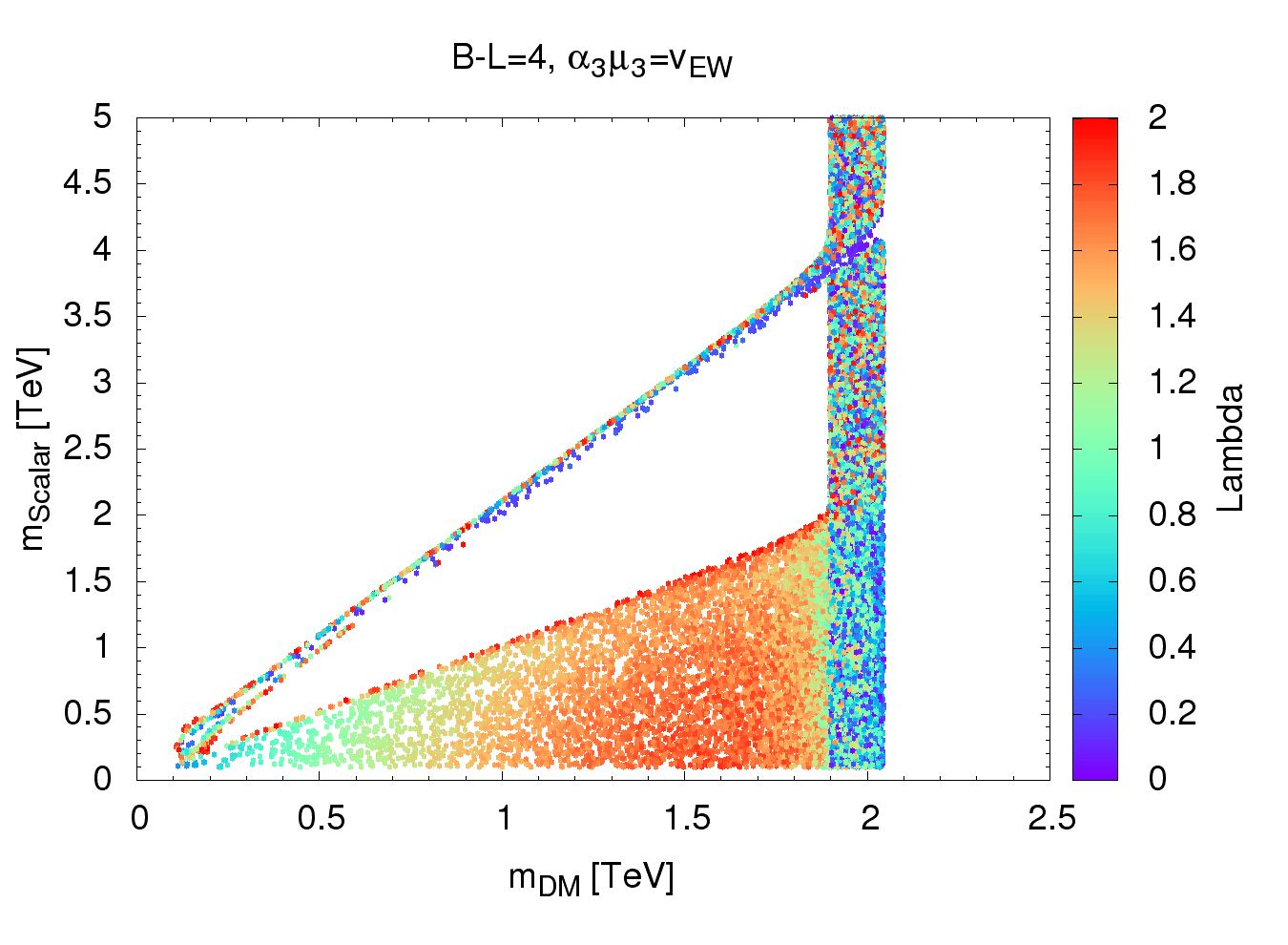}
\includegraphics[width=7.6cm]{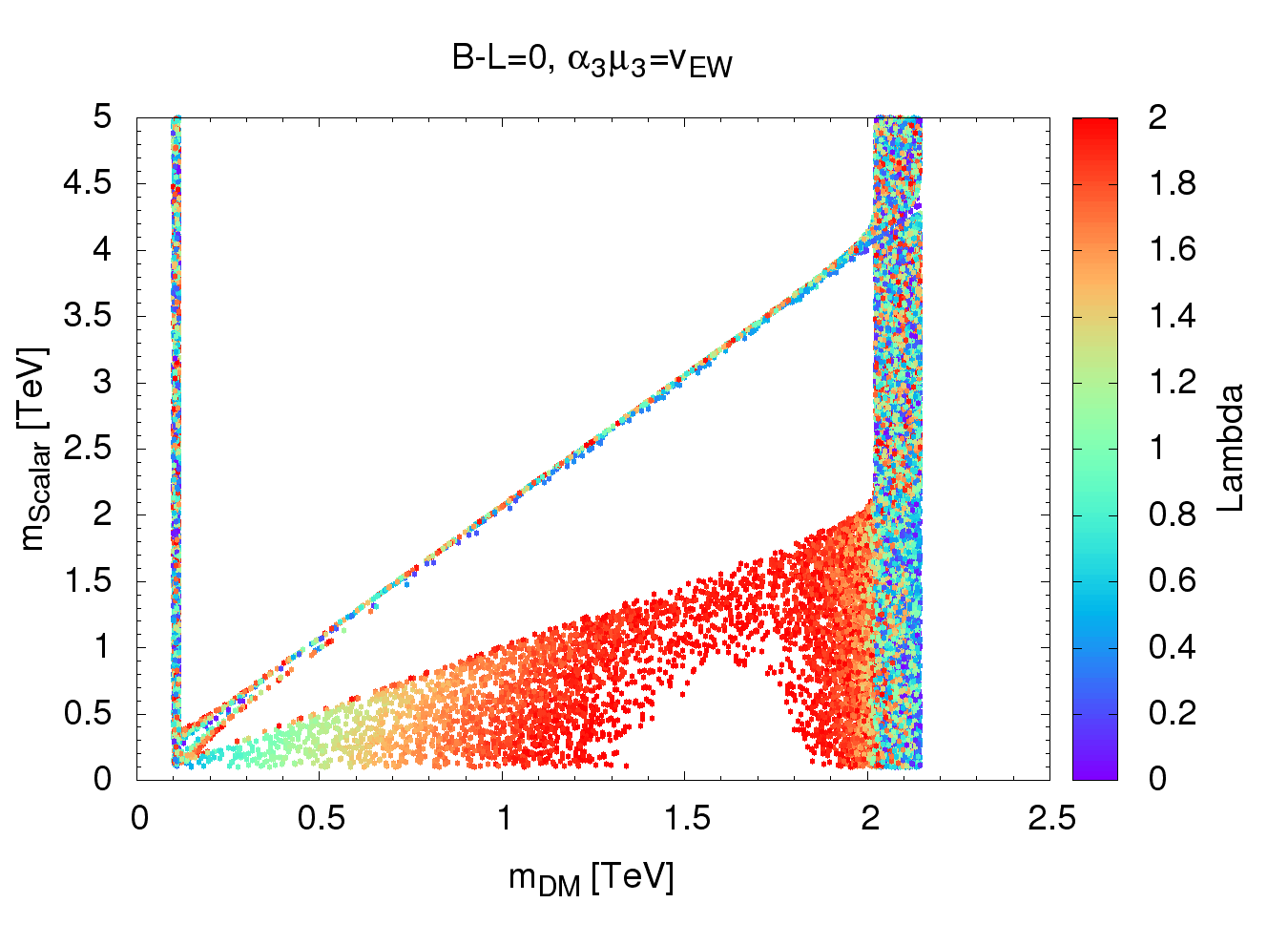}
\includegraphics[width=7.6cm]{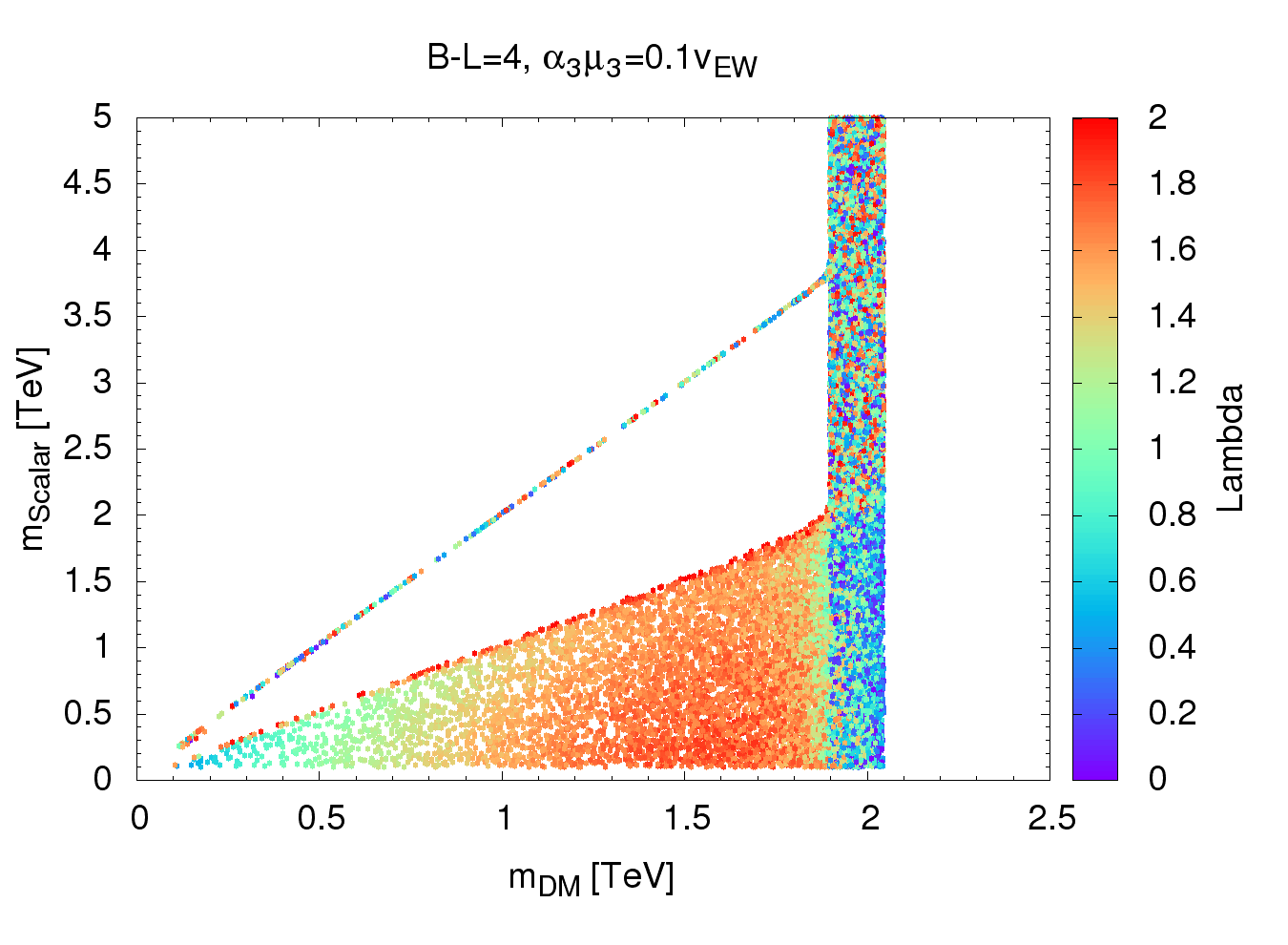}
\includegraphics[width=7.6cm]{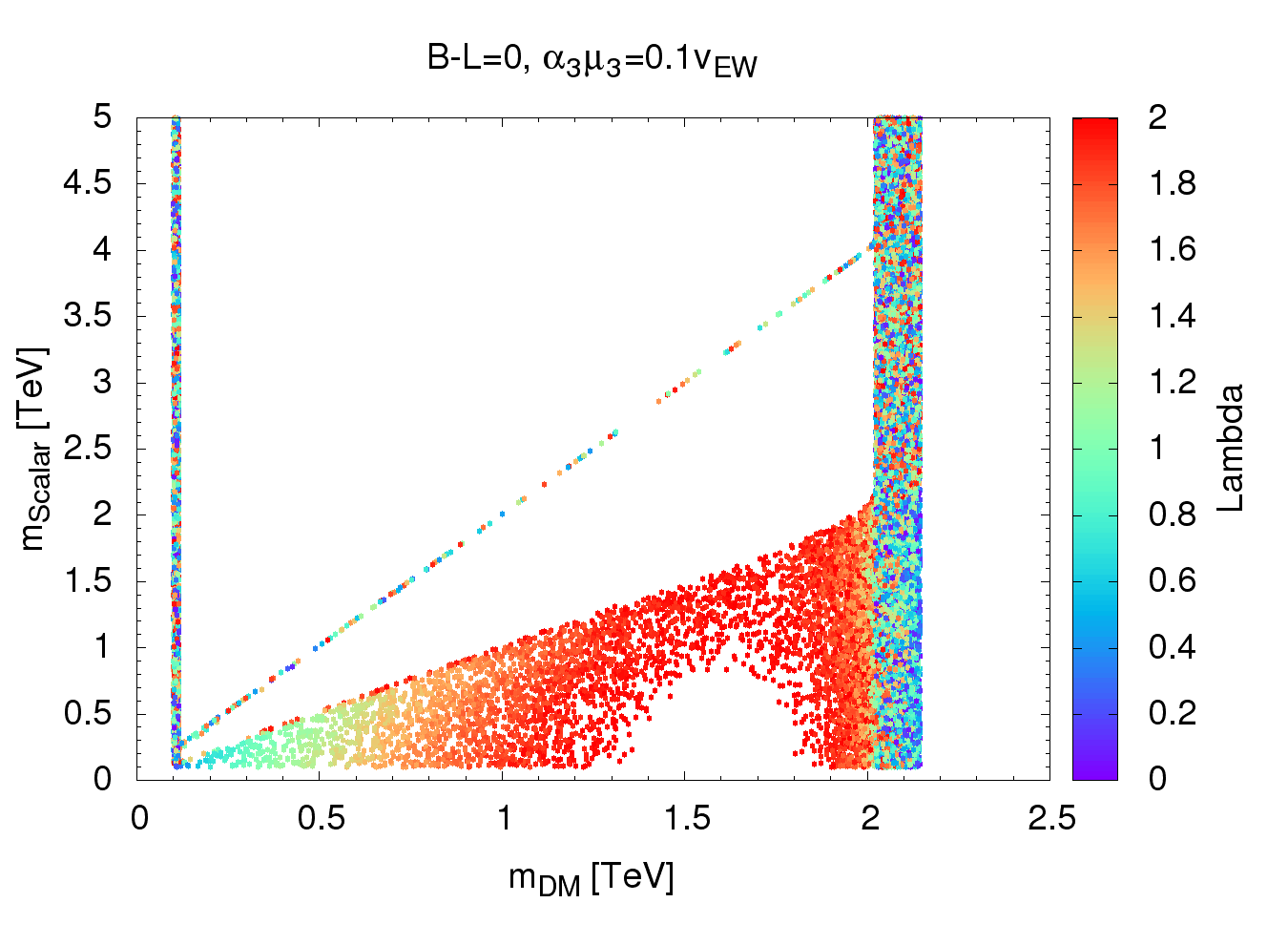}
\mycaption{Scatter plots in $m_{\rm DM}$ - $m_{\rm scalar}$ plane
showing the allowed parameter space satisfying the relic density.
We vary $\lambda$ in the range 0 to 2 in each panel.
The left panel is for $B-L = 4$ case, whereas the
right panel is for $B-L = 0$ case. The upper (lower) panels 
are for $\alpha_3\mu_3$ = $v_{\rm EW}$ (0.1 $v_{\rm EW}$).
In all the panels, we take $M_{W_R}=6$ TeV, $M_{Z_R}=7.14$ TeV.}
\label{fig:rdlam2}
\end{figure}
%====================================================

The scatter plots in Fig.~\ref{fig:rdlam2} represent the allowed points which satisfy the experimentally observed relic density as a function of DM mass, $H_1$ mass and $\la$. We have just considered a relatively low DM mass benchmark region with $0.1 {\text{ TeV}} \leq M_{DM} \leq 2.5$ TeV. The left plots are for the $B-L=4$ case while the right pane are for $B-L=0$. The plot for the $B-L=2$ case is very similar to the one for $B-L=0$ and hence has not been included here. To understand this similarity we need to look at the quintuplet spectrum for each of them. For $B-L=4$ there is only one singly charged particle in the quintuplet spectrum while for both $B-L=0,2$ there are two singly charged quintuplet fermions each. The masses of the charged particles are also quite close resulting in a very similar behavior for both these cases at least till the point where the neutral fermion is the lightest. 

Let us first understand the $B-L=4$ case. Looking at the top left plot in Fig~\ref{fig:rdlam2} we clearly see that there are three well-defined distinct regions. A narrow straight line with $M_{DM} \approx M_{H_1}/2$, a triangular region bounded from above by a straight line $M_{DM}=M_{H_1}$ and a rectangular region around $1.9{\text{ TeV}} \lesssim M_{DM} \lesssim 2.05$ TeV. The narrow straight line corresponds to the s-channel annihilation of two DM particles through a $H_1$ boson. There are actually two lines here with two points for each $H_1$ mass. If we look back at Fig.~\ref{fig:rd4} we see that at a DM mass of $M_{H_1}/2$ there is a sharp dip with two points satisfying the correct relic density, one before and another after the resonance point. Similarly for any scalar mass there should be two such points and hence two narrow straight lines. 
The triangular region is the one corresponding to the t-channel annihilation of two DM particles into two $H_1$ bosons. A close inspection of this region will show that the upper boundary of the plot is lined by points which are of $\la\sim 2$. These are the points where the DM mass is just equal to the scalar mass and the annihilation is only possible for very large cross-section due to the phase space suppression. All the underlying points are where $M_{DM} > M_{H_1}$. In this region there is a monotonic increase in $\la$ as we move from low to high DM mass (for a fixed $M_{H_1}$). Since this process is t-channel, the annihilation cross-section would decrease an the mass difference $M_{DM}-M_{H_1}$ increases and a larger value of $\la$ is needed to compensate for this decrease. The value of $\la$ in this region should also increase as we move downward towards a lower $M_{H_1}$ for a constant DM mass for the exact same reason. 

The rectangular region around 2 TeV is due to the co-annihilation of the DM with a charged particle through a $W_R$ boson. This region should be independent of $\la$ but the plot shows something completely different. We see that in the parts of this region overlapping with the other two, only small values of $\la$ are allowed. Actually in the overlapping regions there are two processes contributing to the decrease in relic density and if the $W_R$ co-annihilation process has to dominate, the other two processes which are both proportional to some power of $\la$ ($\la^4$ and $\la^2$ for the t- and s-channel processes respectively) should be small. Hence the low $\la$ points are only allowed in these regions. 

The $B-L=2$ plot in Fig.~\ref{fig:rdlam2} is similar in nature to the previous case except there is a new region here which is a line parallel to the Y-axis around a DM mass of 200 GeV. If we look back at Fig.~\ref{fig:rd2} we see that the relic density is initially increasing and crosses the experimentally observed line at around the 200 GeV DM mass. This point is independent of $\la$ and gives rise to the vertical line here. Another important new observation here is that now a part of the triangular region can never satisfy the relic density constraints irrespective of the value of $\alpha_3 \mu_3$. As has been discussed earlier, moving from low to high masses in the triangular region requires the annihilation cross-section to progressively increase as well. Thus we require a larger value of $\la$ but since we only allow $\la\leq 2$ there are some parts which simply cannot produce enough annihilation and the relic density is always higher that the observed limit. This situation is remedied as we move closer to the $W_R$ mediated co-annihilation region as now both the t-channel annihilation and the s-channel co-annihilation together can contribute to decrease the relic density to the correct experimental limit. Independently just by increasing the value of $\la$ to include points upto $\la<3$ will result in complete disappearance of this empty patch. The relic density constraints can then be satisfied over the entire parameter region considered here. 

%%======================
%\begin{figure}[ht!]
%\centering
%\includegraphics[width=3.3in]{RD_bml4al01.pdf}
%\includegraphics[width=3.3in]{RD_bml2al01.pdf}
%%\includegraphics[width=2.3in]{MD4.pdf}
%\caption{{Relic density plot a for lower value of $\alpha_3 \mu_3$ = 0.1 $v_{EW}$.}}
%\label{fig:al01}
%%\label{fig:rdlam35}
%\end{figure}
%%=======================

The case with a smaller $\alpha_3 \mu_3$ = 0.1 $v_{EW}$ are plotted in the lower panel of Fig.~\ref{fig:rdlam2}. The only difference compared to the upper panel plots (with $\alpha_3 \mu_3 =v_{EW}$) is that the narrow straight line here is indeed one line instead of two. This, from our earlier observation, is due to the much narrower dip in the s-channel scalar annihilation region for a smaller value of trilinear coupling.

%For completeness we have included a plot of $B-L=2$ in Fig.~\ref{fig:rdlam35} with $\la \leq 3.5$ which is approaching the perturbative limit for the coupling\footnote{Perturbativity requires $\la<\sqrt{4\pi}$.}. Here there is no such empty region as now with an increased coupling (and hence increased annihilation cross-section) it is possible to get the correct relic density for the entire region.

%\begin{figure}[ht!]
%\centering
%\includegraphics[width=3.3in]{RD_bml2al1l35.pdf}
%\includegraphics[width=3.3in]{correct_RD3.pdf}
%\includegraphics[width=2.3in]{MD4.pdf}
%\caption{{Relic density plot for a larger range of $\lambda$.}}
%\label{fig:al01}
%\label{fig:rdlam35}
%\end{figure}

\subsection{Direct Detection}
\label{DD}

\begin{figure}[htb!] \begin{center} \includegraphics[width=0.75\textwidth]{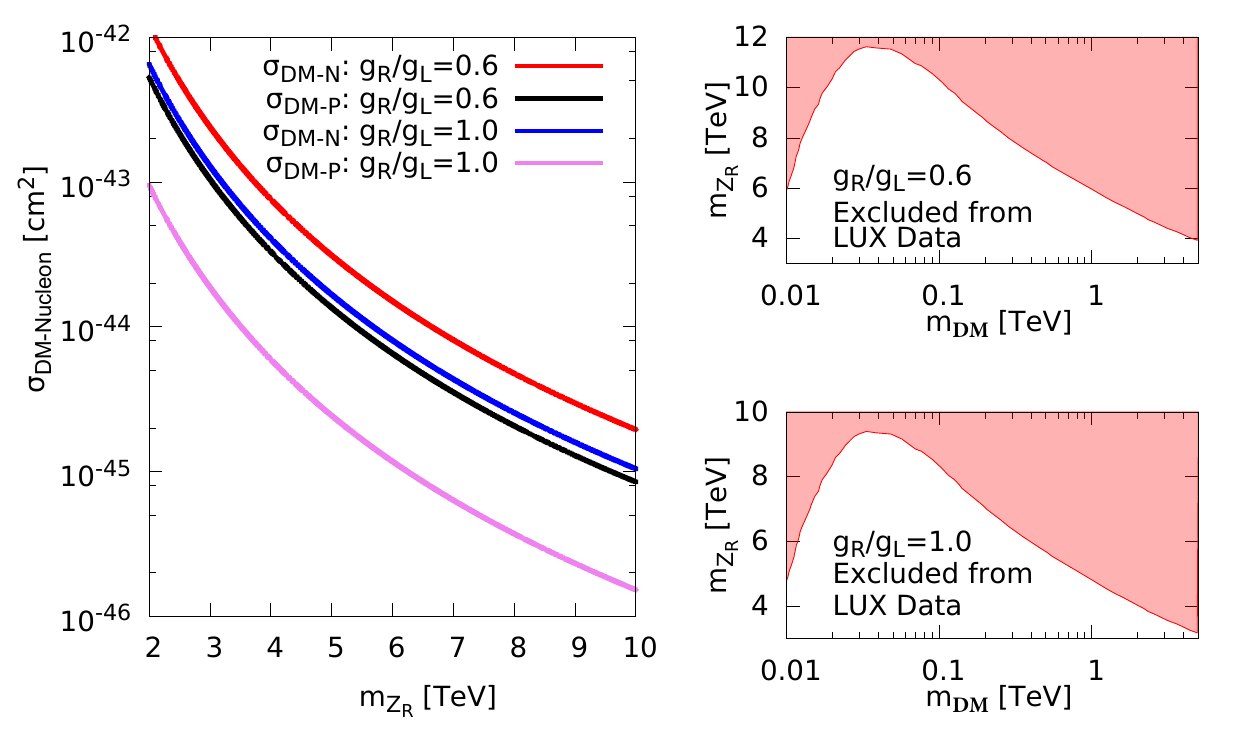} \mycaption{$\chi^0$-proton and $\chi^0$-neutron scattering cross-sections are shown as a function of $m_{Z_R}$ considering two different values of $g_R/g_L$ = 0.6 and 1.0 in the left panel. The top right panel depicts the colored region in $m_{DM}$--$m_{Z_R}$ plane which satisfies the LUX~\cite{Carmona-Benitez:2016byn} upper bound on DM-nucleon scattering cross-section for $g_R/g_L$ = 0.6. We show the same for $g_R/g_L$ = 1.0 in the bottom right panel.} \label{DM} \end{center} \end{figure}

This model can lead to quite significant DM-nucleon scattering cross-section via the $Z_R$-boson exchange diagram, resulting in stringent constraints from DM direct detection experiments. This constraint would be most severe for higher $B-L$ cases while for $B-L$ = 0, the $\chi^0-Z_R$ interaction itself is absent resulting in no significant bounds in this case. We thus study the case of maximal $B-L$(= 4) where the DM-nucleon scattering, mediated by $Z_R$, would be suppressed  by $1/M_{Z_R}^4$. The left panel of Fig.~\ref{DM} gives the scattering cross-section of $\chi_0$-proton and the $\chi_0$-neutron as a function of $m_{Z_R}$ for two different values of $g_R/g_L$. A smaller value of $g_R/g_L$ leads to a larger cross-section and hence requires a larger $Z_R$ mass to evade the direct detection limits. This fact can be easily seen in the right panel of Fig.~\ref{DM} where we have plotted the direct detection bound from LUX~\cite{Carmona-Benitez:2016byn} in the $m_{DM}$--$m_{Z_R}$ plane. The shaded region is consistent with the LUX data and hence a $Z_R$ mass greater than 7 TeV is safe for a DM mass above 100 GeV for $g_R/g_L$ = 1 as has been chosen throughout this paper. 

%================================================================

%\begin{figure}[htpb!]
%\centering
%\includegraphics[width=12.0 cm, height=7.0cm]{DM.pdf}
%\mycaption{$\chi^0$-proton and $\chi^0$-neutron scattering cross-sections are shown as a function of $m_{Z_R}$ considering two different values of $g_R/g_L$ = 0.6 and 1.0 in the left panel. The top right panel depicts the colored region in $m_{DM}$--$m_{Z_R}$ plane which satisfies the LUX~\cite{Carmona-Benitez:2016byn} upper bound on DM-nucleon scattering cross-section for $g_R/g_L$ = 0.6. We show the same for $g_R/g_L$ = 1.0 in the bottom right panel. }
%\label{DM}
%\end{figure}
%================================================================

\section{Phenomenology of the Singlet Scalar}
\label{singlet}
 
\begin{figure}[ht!]
\includegraphics[width=7.75 cm]{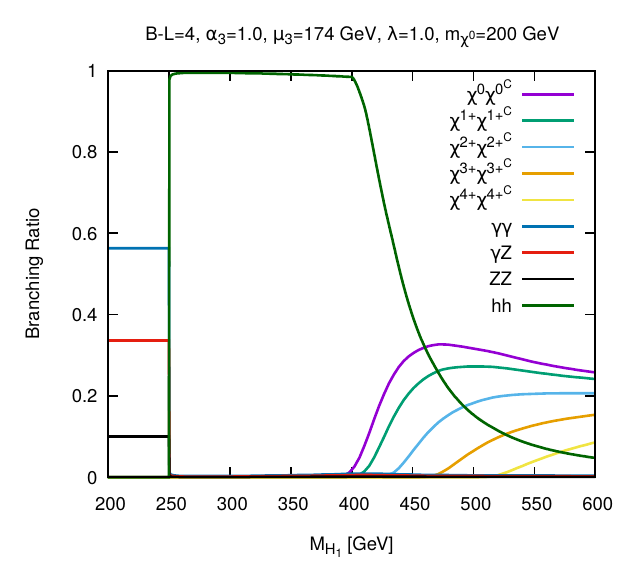}
\includegraphics[width=7.75 cm]{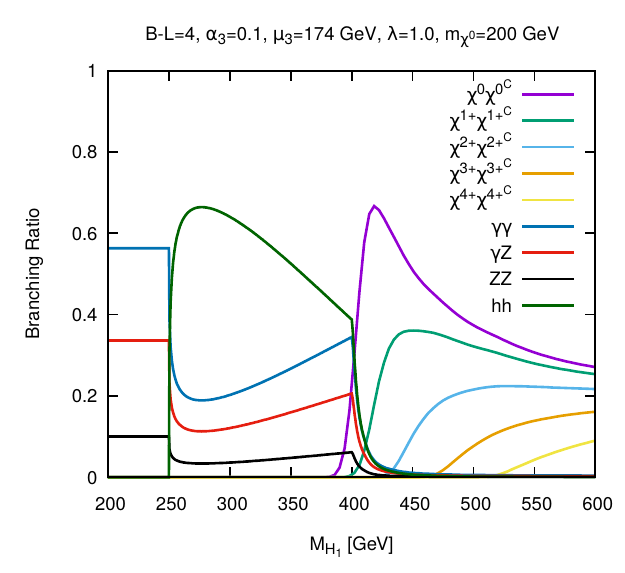}
\mycaption{The decay branching ratios of singlet scalar ($H_1$) as a function of its mass. The left (right) panel is for $\alpha_3\mu_3 = v_{EW}~(0.1v_{EW})$. In both the panels we consider $B-L=4$ case with $\la=1.0$ and $m_{\chi^0}=200$ GeV.}
\label{fig:br}
\end{figure}

Although the singlet scalar was introduced to satisfy the dark matter relic density for almost any DM mass compared to only a few points in its absence, it gives rise to interesting signatures at the collider experiments. before going into the details of production cross-section and collider signature it is important to study the decays of singlet scalar ($H_1$). At tree level, $H_1$ couples with a pair of SM Higgs bosons or with a pair of quintuplet fermions. Therefore, if kinematically allowed, $H_1$ dominantly decays into a pair of Higgses of a pair of quintuplet fermions. $H_1$ also has loop induced couplings with a pair of photons, $Z$-bosons and photon-$Z$ pair. The coupling of both photon and $Z$-boson with the quintuplet fermions being proportional to the electric charge of the fermions (see Eq.~\ref{eq:gaudm}), the loop induced decays can be quite significant as it involves the multi-charged quintuplet fermions running in the loop. In particular, the diphoton decay could be as significant as other decay modes in certain parts of the parameter space. The loop induced interactions (in particular interaction with a pair of photons) play the most crucial role in the production and phenomenology of $H_1$ in the context of hadron collider experiments. Since the $B-L$ = 4 quintuplet would contain fermions of highest charge multiplicity, we will only consider this case for our analysis in this section as it will lead to the strongest constraints on the model. The decay width of the singlet scalar in various decay channels are given as 
\begin{eqnarray}
\Gamma_{hh}&=&\frac{\alpha_3^2 \mu_3^2}{8 \pi m_{H_1}}\left(1-\frac{4 m_{h}^2}{m_{H_1}^2}\right)^{\frac{1}{2}},\nonumber\\
\Gamma_{\chi^i \bar \chi^i}&=&\frac{\lambda^2}{8\pi}m_{H_1}\left(1-\frac{4 M_{\chi^i}^2}{m_{H_1}^2}\right)^{\frac{3}{2}},\nonumber\\
\Gamma_{\gamma\gamma}&=&\frac{\alpha^2 m_{H_1}^3 \lambda^2}{256 \pi^3}\left|\sum_{\chi^i} \frac{Q^2_{\chi^i}}{M_{\chi^i}}A_{\frac{1}{2}}\left(\frac{m_{H_1}^2}{4M_{\chi^i}^2}\right)\right|^2,\nonumber\\
\Gamma_{Z\gamma}&=&\frac{\alpha^2 m_{H_1}^3 \lambda^2}{128 \pi^3}{\rm tan}^2\theta_W\left|\sum_{\chi^i} \frac{Q^2_{\chi^i}}{M_{\chi^i}}A_{\frac{1}{2}}\left(\frac{m_{H_1}^2}{4M_{\chi^i}^2}\right)\right|^2,\nonumber\\
\Gamma_{ZZ}&=&\frac{\alpha^2 m_{H_1}^3 \lambda^2}{256 \pi^3}{\rm tan}^4\theta_W\left|\sum_{\chi^i} \frac{Q^2_{\chi^i}}{M_{\chi^i}}A_{\frac{1}{2}}\left(\frac{m_{H_1}^2}{4M_{\chi^i}^2}\right)\right|^2,~~~~
\label{decay_width}
\end{eqnarray}
%===================== 
where $\chi^i \subset \{\chi^{++++},~\chi^{+++},~\chi^{++},~\chi^{+}~{\rm and}~\chi^{0}\}$, $M_{\chi^i}$ and $Q_{\chi^i}$ are the mass and charge of the corresponding $\chi^i$ respectively. {The loop function $A_{1/2}(x)$ is given by, $A_{1/2}(x)=2x^{-2}\left[x+(x-1)f(x)\right]$, where, $f(x)=-[{\rm ln}\{(1+\sqrt{1-x})/(1-\sqrt{1-x})\}-i\pi]^2/4$ for $x >1$ and $f(x)={\rm arcsin}^2{\sqrt x}$ for $x \le 1$.} In Fig.~\ref{fig:br} we have plotted the scalar decay branching ratios as a function of the singlet mass for a fixed value of DM mass, $\la$ and for two different values of $\alpha_3$. The left panel corresponds to $\alpha_3$ = 1.0 while the right panel is for $\alpha_3$ = 0.1. {For $M_{H_1}<250$ GeV, the di-Higgs decay is kinematically forbidden and hence, the only possible decay modes are the loop induced decays into a pair of SM gauge bosons. The decay into a pair of $W^\pm$-bosons are highly suppressed by the small $W_L$--$W_R$ mixing and hence, not shown in Fig.~\ref{fig:br}. The di-Higgs decay mode becomes kinematically allowed for $M_{H_1}>250$ GeV for a 125 GeV SM Higgs. In this region of the parameter space, the branching ratios depends on two parameters, namely the Yukawa coupling $\lambda$ (which determines the strength of the loop induced diboson-singlet scalar interactions) and $\alpha_3$ (which determines the di-Higgs decay width) in the scalar potential.} We clearly see that as we decrease the value of $\alpha_3$, the diphoton branching ratio increases compared to the di-Higgs. A similar phenomenon will also take place if one increases the value of $\la$ keeping $\alpha_3$ constant. Once the quintuplet decay channel opens up, the entire decay is almost into the quintuplet fermions with all other channels completely disappearing. It is important to notice the enhancement of the loop induced decay branching ratios into two gauge bosons at around 400 GeV which is the threshold of quintuplet on-shell contribution in the loop.

{The singlet scalar has loop induced coupling with a pair of photons. The production of $H_1$ at the LHC proceeds through photon-fusion process and hence, suppressed by the small parton density of photon inside a proton. In fact, the parton density of photon is so small that most of the older versions of PDF's do not include photon as a parton. However, photo-production is the only way to produce $H_1$ at the LHC. Moreover, if we want to include QED correction to the PDF, inclusion of the photon as a parton with an associated parton distribution function is necessary. And in the era of precision physics at the LHC when PDF's are determined upto NNLO in QCD, NLO QED corrections are important (since $\alpha_{s}^2$ is of the same order of magnitude as $\alpha$) for the consistency of calculations. In view of these facts,  NNPDF \cite{Ball:2014uwa,Ball:2013hta}, MRST \cite{Martin:2004dh} and CTEQ \cite{Schmidt:2015zda} have already included photon PDF into their PDF sets. However, different groups used  different approaches for modeling the photon PDF. For example, the MRST% \cite{Martin:2004dh}
 group used a pasteurization for the photon PDF based on radiation off of primordial up and down quarks, with the photon radiation cut off at low scales by constituent or current quark masses. The CT14QED% \cite{Schmidt:2015zda} 
 variant of this approach constrains the effective mass scale using $ep \rightarrow e \gamma +X$ data, sensitive to the photon in a limited momentum range through the reaction $e\gamma \rightarrow e \gamma$.  The NNPDF group used a more general photon parametrization, which was then constrained by high-energy W, Z and Drell-Yan data at the LHC. For computing the photon-luminosity at the $pp$ collision with 13 TeV center of mass energy, we have used NNPDF23$\char`_$lo$\char`_$as$\char`_$0130 {%\cite{Ball:2014uwa,Ball:2013hta}}
PDF with the factorization scales being chosen to be fixed at $M_{H_1}$. The resonant photo-production cross-section of $H_1$ at the LHC can be computed from its di-photon decay width and LHC photon luminosity at $\sqrt s=M_{H_1}$ as follows: }
\begin{equation}
\sigma(pp\to H_1)~=~\frac{8\pi^2}{s}\frac{\Gamma(H_1\to \gamma \gamma)}{M_{H_1}}\int_{\tau}^1\frac{dx}{x}f_{\gamma}(x)f_{\gamma}(\tau/x),
\end{equation}
{where, $tau=M_{H_1}^2/s$, $f_\gamma(x)$ is the photon PDF and $s$ is the $pp$ center of mass energy. The production cross-section for the heavy singlet is shown in Fig.~\ref{fig:prod} as a function of singlet mass for a few different DM masses. The DM mass is important here because the masses of other quintuplet fermions is determined by the DM mass and radiative corrections. A larger DM mass implies that the masses of the charged states running in the photon fusion loop are also larger and hence a smaller cross-section for singlet production. It is important to notice the bump around $M_{H_1}\sim 2M_{\chi^{4+}}$ due to the threshold enhancement of the diphoton decay width.}

\begin{figure}[ht!]
\centering
\includegraphics[width=12.0 cm, height=7.0cm]{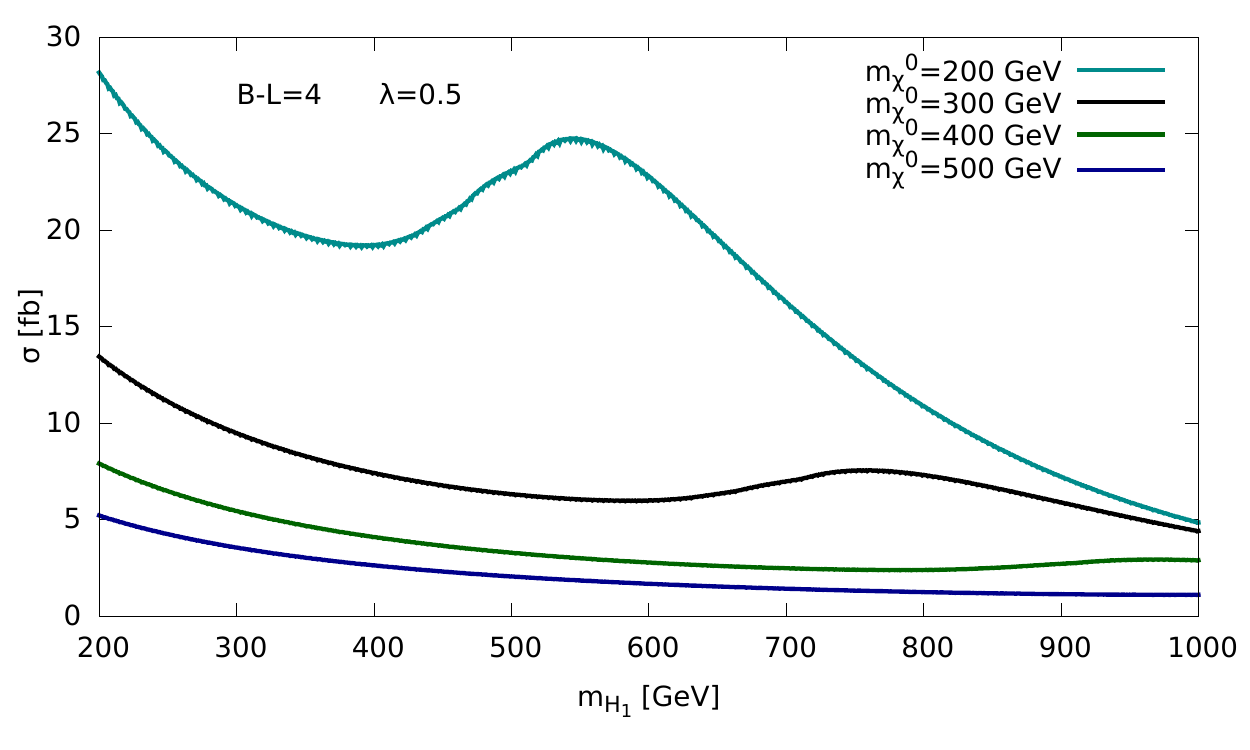}
\mycaption{The production cross-section of singlet scalar ($H_1$) as a function of its mass for various choices of DM mass ($m_{\chi^0}$). The curves are shown for $B-L=4$ case with $\la=0.5$. }
\label{fig:prod}
\end{figure}

\section{Collider Bounds}
\label{bounds}
After production, the singlet scalar dominantly decays into a pair of photons or Higgs bosons as long as the decays into a pair of quintuplet fermions are kinematically forbidden. Therefore, the production and decay of $H_1$ gives rise to interesting resonant diphoton and/or di-Higgs signatures at the LHC. The ATLAS and CMS collaborations of the LHC experiment have already searched for any  new physics signatures in the diphoton \cite{Aaboud:2017yyg} and di-Higgs \cite{Aaboud:2017eta} invariant mass distributions. In absence of any significant deviation from the SM prediction, limits are imposed on the production cross sections times branching ratio of a resonance giving rise to above mentioned signatures. These limits could lead to significant bounds on the DM RD allowed scalar mass {(for example, see Fig.~\ref{fig:rdlam2})}. In our analysis we found that the diphoton bound is a lot more severe especially for the $\alpha_3$ = 0.1 case. In Fig.~\ref{fig:diphot} we have plotted the $H_1$ production cross-section times the diphoton branching ratio ($\sigma_{pp\rightarrow H_1} \times H_1 \rightarrow \gamma \gamma$) and compared with the ATLAS observed limit \cite{Aaboud:2017yyg}. The diphoton-production cross-section is plotted for two different values of DM mass of 200 GeV and 600 GeV for fixed values of $\la$ = 1.0 and $\alpha_3$ = 0.1. 
\begin{figure}[ht!]
\centering
\includegraphics[width=12.0 cm, height=7.0cm]{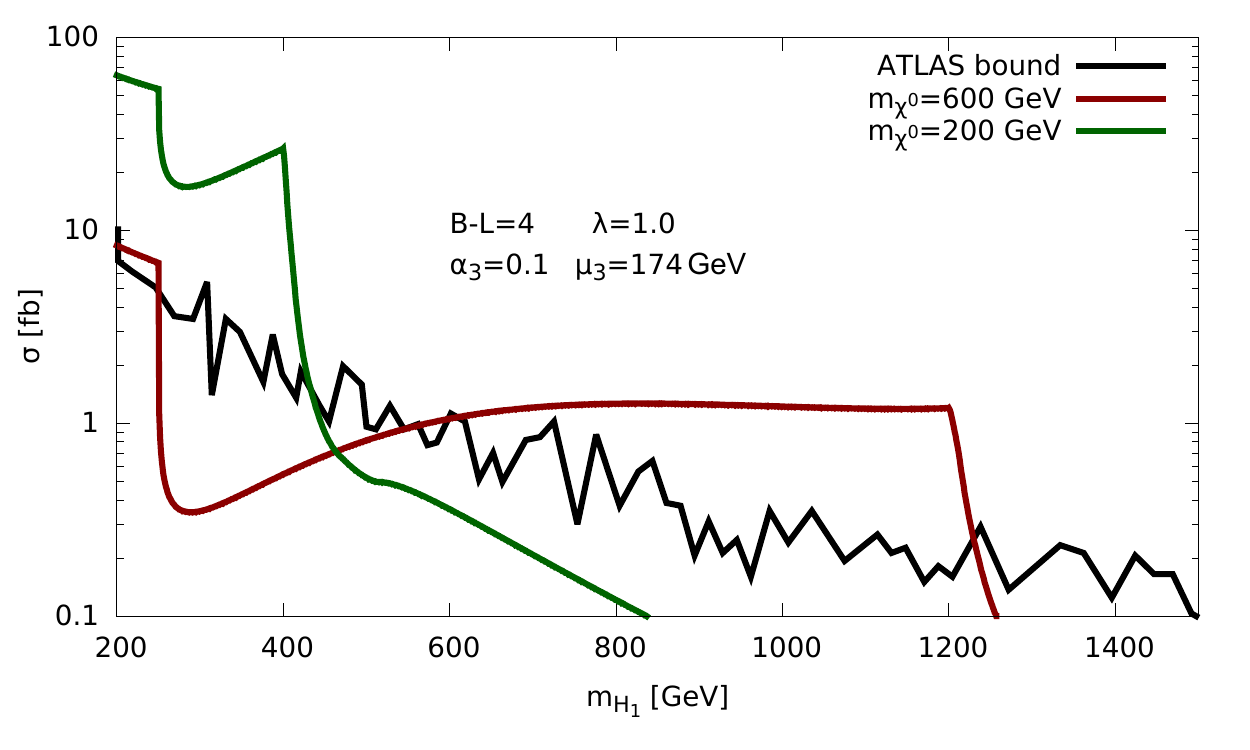}
\mycaption{The diphoton production cross-section is shown as a function of the singlet scalar mass for two different choices of the DM mass. We draw these curves for $B-L$=4 case with $\alpha_3 \mu_3=0.1 v_{EW}$ and $\la=1.0$. For comparison, the solid black line shows the ATLAS bound \cite{Aaboud:2017yyg}.}
\label{fig:diphot}
\end{figure}
For the 200 GeV DM mass, any scalar mass below 430 GeV is ruled out from the experiments. For the 600 GeV DM mass a small region from 630 GeV to 680 GeV along with 730 GeV to 1220 GeV scalar masses would be ruled out. {In is interesting to note that larger values of the scalar masses are excluded for higher DM mass while the smaller values of $M_{H_1}$ remain allowed. This is a consequence of the fact that larger DM mass corresponds to smaller $H_1$ production cross-section and hence, the diphoton signal cross-section  in the smaller $M_{H_1}$ region  are smaller than the ATLAS bound. On the other hand, larger DM mass also corresponds to a  threshold enhancement of the diphoton decay width and hence, diphoton signal rate  at larger $M_{H_1}$. As a result, some part of $M_{H_1}$ region around $M_{H_1}\sim 2M_{\chi^0}$ is excluded for $M_{\chi^0}\sim 600$ GeV.} 

Experimental bounds on the resonant di-Higgs \cite{Aaboud:2017eta} and diphoton \cite{Aaboud:2017yyg} signal cross-sections have a significant impact on the DM allowed regions given in Fig.~\ref{fig:rdlam2}. We have scanned the DM allowed points in Fig.~\ref{fig:rdlam2} to check the consistency of those points with the ATLAS di-Higgs and diphoton searches and the results are presented in Fig.~\ref{fig:DM_bound}. 
\begin{figure}[ht!]
\includegraphics[width=7.75 cm]{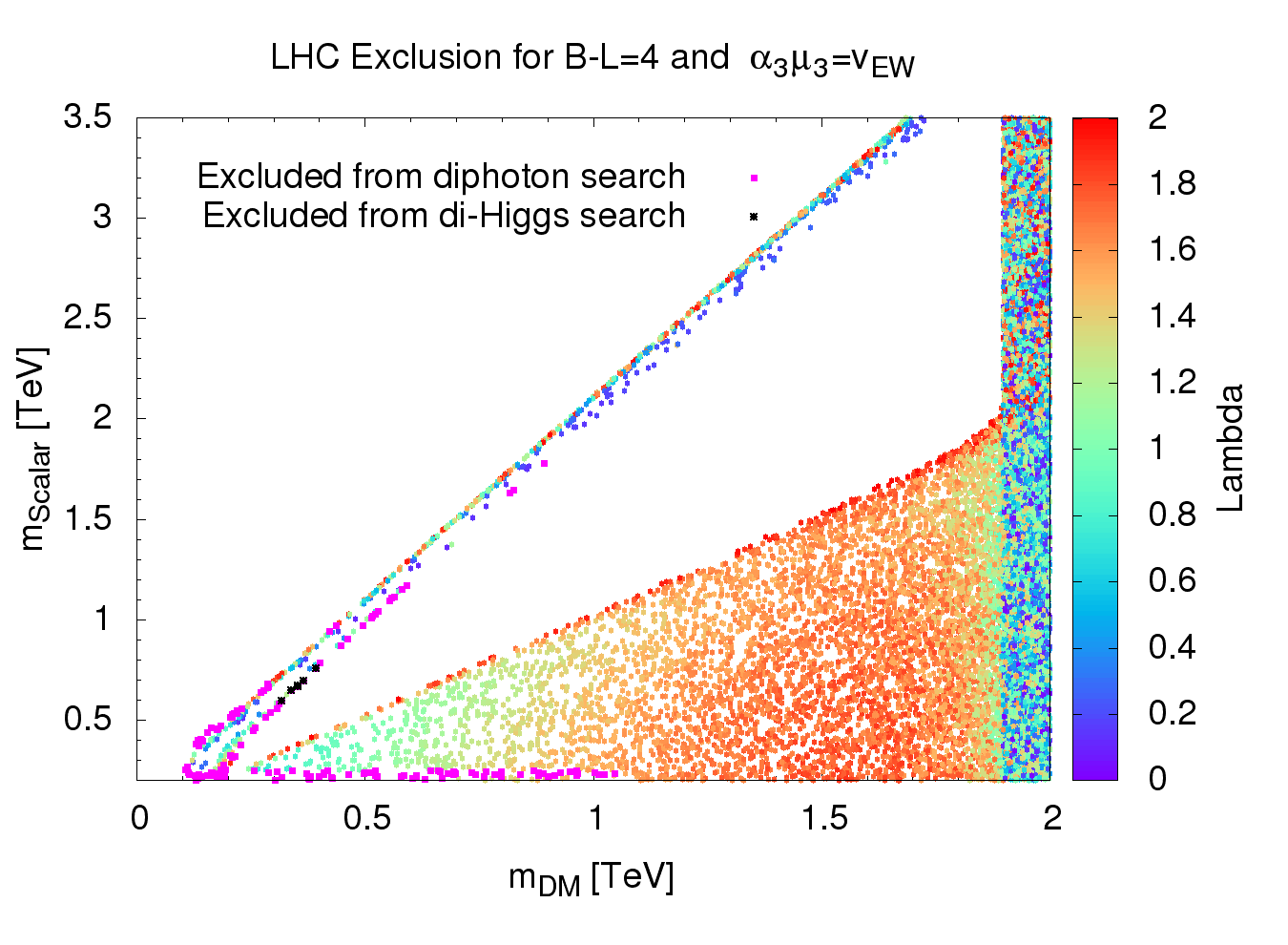}
\includegraphics[width=7.75 cm]{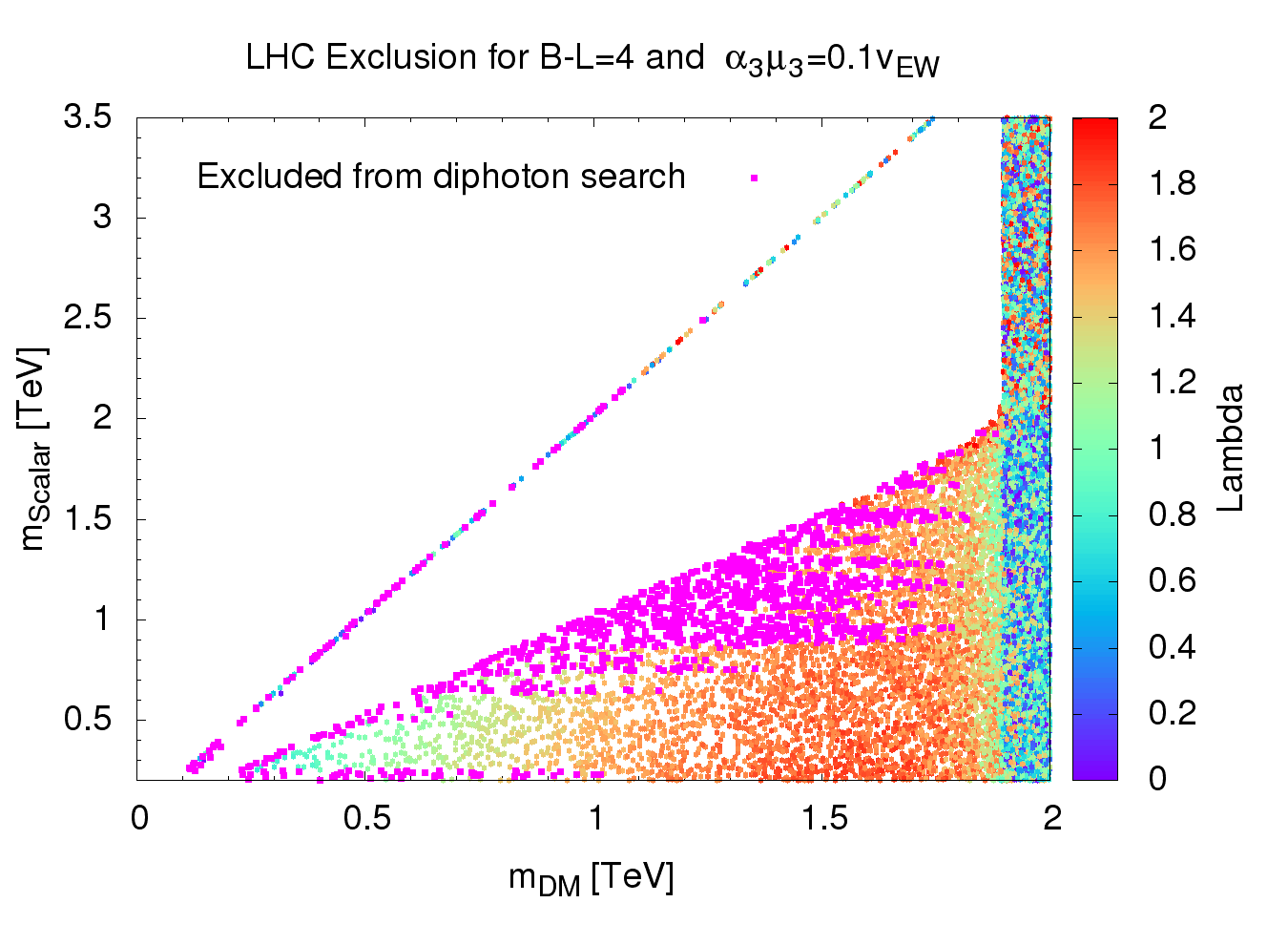}
\mycaption{The allowed parameter space in $m_{DM}-m_{scalar}$ plane satisfying the relic density for $B-L=4$ case. The left (right) panel is for $\alpha_3 \mu_3 = v_{EW} (0.1 v_{EW})$. In both the panels we vary $\la$ in the range 0 to 2. The pink (black) points show the regions excluded from diphoton \cite{Aaboud:2017yyg} (di-higgs \cite{Aaboud:2017eta}) search at LHC. }
\label{fig:DM_bound}
\end{figure}
The pink points in the plots are ruled out from diphoton search while the black points are ruled out by di-Higgs searches. As one would expect, the di-Higgs bounds are only applicable for case with $\alpha_3$ = 1.0 as the two Higgs final state scalar decay branching ratio is quite large in this case. The diphoton bounds are much stronger for the lower $\alpha_3$ as the diphoton decay branching ratio is much larger here. Even though a part of the parameter space is ruled out a large part of it is still remains which can explain all the observations from both the collider and dark matter experiments.

\section{Summary and Conclusions}
\label{conclusion}

To summarize, we have performed the dark matter and collider phenomenology of a left-right symmetric ($SU(3)_C\times SU(2)_L\times SU(2)_R\times U(1)_{B-L}$ gauge symmetry) model with an additional $SU(2)_R$ quintuplet fermion and a singlet scalar. The motivation for introducing the quintuplet fermion is to obtain a viable candidate for cold dark matter. The neutral component of the quintuplet fermion, being weakly interacting and stable (if lightest among the other components of the quintuplet), could be a good candidate for dark matter. The dark matter, in this model, can interact with ordinary matter via the exchange of a $SU(2)_R$ gauge boson (in particular, $Z_R$). The bounds on the dark matter-nucleon scattering cross-sections from the direct dark matter detection experiments such as LUX exclude $M_{Z_R}$ below few TeV for a sub-TeV dark matter. Moreover, the gauge interactions of the neutral quintuplet fermion with massive ($>$ few TeV) $SU(2)_R$ gauge bosons result into small annihilation and co-annihilation cross-sections and thus, predict relic density which is much larger than the observed WMAP/PLANCK results. The observed relic density can only be satisfied for few discrete values of the dark matter mass near $W_R$/$Z_R$ resonance region (near $M_{W_R}/2$ and $M_{Z_R}/2$). Therefore, in the framework of left-right symmetry with a quintuplet dark matter candidate, sub-TeV dark matter masses are ruled out from the direct detection experiments and relic density constraints. Moreover, an experimentally consistent dark matter candidate in the range of few TeV is only possible for $B-L=4$ case. To resolve these issues we introduce a singlet scalar in the above mentioned framework. The Yukawa coupling of the singlet scalar with the quintuplet fermion gives rise to a host of new annihilation channels for the Dark matter. We perform in detail the dark matter phenomenology in this singlet scalar extended scenario. We show that the WMAP/PLANCK measured dark matter relic density can be satisfied over a large range of dark matter masses including sub-TeV range. Moreover, the neutral component of the quintuplet fermion with $B-L=2$ and $0$ cases can give rise to an experimentally consistent candidate for dark matter as long as they are the lightest member of the quintuplet. 

We also study the collider signatures of the singlet scalar in details. Being singlet, it has no tree level interactions with the SM gauge bosons or Yukawa interactions with the ordinary leptons and quarks. However, the Yukawa interaction with the quintuplet fermion is allowed by the gauge symmetry. The interactions of the singlet scalar with a pair of EW gauge bosons arise from the loop induced higher dimensional operators. On the other hand, the scalar potential contains the interactions involving the singlet scalar and a pair of SM Higgs bosons. It enables us to study the loop induced ($\gamma\gamma$, $ZZ$ and $Z\gamma$) as well as tree level decays ($hh$ and a pair of quintuplet fermions) of the singlet scalar. We find that as long as the decay of the scalar to a pair of quintuplet fermions are kinematically forbidden, it only decays to a pair of Higgs bosons or a pair of photons with significant branching ratios. In view of this, we study the photo-production (photon-photon fusion process) of the singlet scalar at the LHC with 13 TeV center-of-mass energy. The photo-production of the singlet scalar and its subsequent decay give rise to interesting resonant diphoton and di-Higgs signatures at the LHC. New physics contributions to the resonant diphoton and di-Higgs productions have already been studied in detail by the ATLAS and CMS Collaborations. We use the most recent ATLAS bounds on the resonant di-Higgs and diphoton cross-sections to show that some part of the dark matter relic density allowed parameter space in our model could be ruled out. It is worthwhile to mention that a significant part of the parameter space in our model is still consistent with the dark matter direct detection constraints, WMAP/PLANCK results as well as the LHC bounds. Future LHC data will be able to probe this part of the parameter space which is still allowed. Therefore, the singlet scalar extended quintuplet MDM left-right symmetric model will be able to explain any future LHC excesses either in the resonant di-Higgs or diphoton channels. On the other hand, the absence of any such excesses could lead to more stringent bounds on the parameter space of our model. 

%===================
\section{Acknowledgment}
%===================

S.K.A. acknowledges the support from the DST/INSPIRE Research Grant [IFA-PH-12], 
Department of Science and Technology, India. K.G. is supported by the DST/INSPIRE
Research Grant [DST/INSPIRE/04/2014/002158]. A.P. is supported by the SERB 
National Postdoctoral fellowship [PDF/2016/000202]. A.P. thanks the organizers of ``Candles of Darkness" conference held at ICTS, Bangalore, India during June, 2017 for giving the opportunity to present the preliminary results of this work.

%%======
%\appendix
%%======

%%============================================
%%=============================================

%=========================
\bibliographystyle{JHEP}
\bibliography{collider}
%==========================

\end{document}